\documentclass[fleqn,usenatbib]{mnras}

\usepackage{newtxtext,newtxmath}
\usepackage[T1]{fontenc}

\DeclareRobustCommand{\VAN}[3]{#2}
\let\VANthebibliography\thebibliography
\def\thebibliography{\DeclareRobustCommand{\VAN}[3]{##3}\VANthebibliography}

\usepackage{graphicx}	% Including figure files
\usepackage{amsmath}	% Advanced maths commands
\usepackage{subcaption}
\usepackage{xcolor}
\usepackage{comment}
\usepackage{wrapfig,lipsum,booktabs}

\newcommand{\di}[1]{{\textcolor{black}{#1}}}

\def\gtsim {>\kern-1.2em\lower1.1ex\hbox{$\sim$}~}   % Greater than sim
\def\ltsim {<\kern-1.2em\lower1.1ex\hbox{$\sim$}~}   % Less than sim

\title[The impact of supernova feedback on MZRs]{The impact of supernova feedback on the mass--metallicity relations}

\author[Ibrahim \& Kobayashi]{
Dyna Ibrahim$^{1}$\thanks{E-mail: d.ibrahim3@herts.ac.uk}
and Chiaki Kobayashi$^{1}$
\\
$^{1}$Centre for Astrophysics Research, Department of Physics, Astronomy and Mathematics University of Hertfordshire, College Lane, Hatfield AL10 9AB, UK
}

\date{Received 12th July 2023}

\pubyear{2023}

\begin{document}
\label{firstpage}
\pagerange{\pageref{firstpage}--\pageref{lastpage}}
\maketitle

% Abstract of the paper
\begin{abstract}
Metallicity is a fundamental physical property that strongly constrains galaxy formation and evolution. The formation of stars in galaxies is suppressed by the energy released from supernova explosions and can be enhanced by metal production. In order to understand the impact of this supernova feedback, we compare four different feedback methods, ejecting energy in thermal, kinetic, stochastic and mechanical forms, into our self-consistent cosmological chemodynamical simulations. To minimise other uncertainties, we use the latest nucleosynthesis yields that can reproduce the observed elemental abundances of stars in the Milky Way. For each method, we predict the evolution of stellar and gas-phase metallicities as a function of galaxy mass, i.e., the mass--metallicity relations. We then find that the mechanical feedback can give the best match to a number of observations up to redshift $z\sim3$, although the predicted gas-phase metallicities seem to be higher than observed at $z\gtsim 1$. The feedback modelling can be further constrained by the metallicities in distant galaxies with the James Web Space Telescope and those of a large sample with ongoing and future spectroscopic surveys.
\end{abstract}

% Select between one and six entries from the list of approved keywords.
% Don't make up new ones.

%%%%%%%%%%%%%%%%%%%%%%%%%%%%%%%%%%%%%%%%%%%%%%%%%%
%\input{section1/Introduction}
%\input{section2/model}

%%%%%%%%%%%%%%%%% BODY OF PAPER %%%%%%%%%%%%%%%%%%

\section{Introduction}

The evolution of elemental abundances in the universe across cosmic time is essential to understand the formation and evolution of galaxies \citep[e.g.,][for a review]{kobayashi_taylor_2023}. While the evolution of dark matter in the standard $\Lambda$ cold dark matter ($\Lambda$-CDM) cosmology is well understood, one of today's greatest challenges is understanding the evolution of baryonic matter from primordial elements produced in the Big Bang nucleosynthesis to elements heavier than helium produced in stars. Metals are observed in the local to the distant galaxies.
The abundances of metals in galaxies give information about the star formation rate (SFR), gas outflows, and inflow during the galaxies' histories. The study of metallicity also provides crucial information about %star formation history and 
the exchange of metals between stars, the cold interstellar gas, and the diffuse surrounding gas.

Understanding the origin and behaviour of elements is subject to several studies. Chemical elements are produced during different astronomical events. Hydrogen and helium form through the Big Bang nucleosynthesis, while carbon and heavier elements form in stellar nucleosynthesis from core-collapse supernovae (SNe), asymptotic giant branch (AGB) stars, thermonuclear explosions observed as Type Ia supernovae (SNe Ia), and neutron star mergers observed as kilonovae \citep{Kobayashi_Karakas_Lugaro2020}. Metals are produced in stars and are ejected into the interstellar medium (ISM), circumgalactic medium (CGM), and the intergalactic medium (IGM) \citep[e.g.,][]{perou2020ARAA}. This ejection happens through losing the outer gaseous envelopes of old/dying stars or the explosion of massive stars as supernovae (with initial masses $\gtsim10M_\odot$). The energy released through stellar winds and supernovae explosions is known as stellar feedback \citep[e.g.,][]{Larson1974}. Feedback can efficiently suppress star formation by heating and evaporating dense, star-forming clouds, generating turbulent supersonic shocks, and generating outflows that eject gas from the galaxy. At low halo masses, the dominant feedback is from massive stars (stellar winds, supernova explosions, photoionization, and radiation pressure). Whereas at higher masses, active galactic nucleus (AGN) feedback dominates (e.g.,  \citealt{Silk2013, Taylor2015}). 
Different feedback methods are used in different cosmological simulations, such as thermal feedback (e.g., \citealt{Katz1992}) %(TreeSpH), 
kinetic feedback \citep{Navarro_White_1993},
stochastic feedback \citep{Dalla2012},  and mechanical feedback \citep{Hopkins2018, Smith2018}. 
%\ck{Hopkins18?}
%\ck{
In this paper we investigate the impact of supernova feedback on the metallicities of galaxies using the same stellar yields in our cosmological simulations.
%\ck{[This paragraph is good for your thesis, but shorten in this paper. Make a few sentences to connect to the next paragraph.]}

Cosmological simulations consider two different processes for the evolution of galaxies over cosmic time: (1) The hierarchical growth of dark-matter structures on timescales proportional to redshift \citep{Press1974}. (2) The baryonic physics on timescales that are impacted by the processes such as radiative cooling, star formation, and feedback \citep{White_Rees1978}. The simulation of galaxy formation and evolution remains a significant challenge as it extends from large-scale structures along dark matter filaments to star formation scales. Assumptions and approximations are therefore necessary and depend on the scale we want to resolve. For instance, the semi-analytic models (SAMs) \citep{White_Frenk1991} compute the baryonic physics separately from the dark matter. SAMs treat each galaxy as an unresolved object and provide a statistical sample of galaxies. On the other hand, hydrodynamical simulations model the hydrodynamics and gravitational laws and can simulate the baryonic physics simultaneously with the dark matter self-consistently. These simulations can also predict the internal structure of galaxies (i.e., kinematics and spatial distributions). However, the results are still limited to a finite resolution. Therefore, all currently available hydrodynamical models implement %`sub-grid'
analytical laws to attempt to capture the effects of the above mentioned sub-galactic processes on a galaxy scale.

Several hydrodynamical simulations are used to predict the %chemical 
evolution in galaxies, with very different input physics in each simulation. For example, the EAGLE simulations \citep{Schaye2015} use stochastic feedback \citep{Dalla2012}. Illustris uses bipolar winds, IllustrisTNG \citep{Pillepich2018} uses isotropic, kinetic (wind) feedback from \citep{Springel2003}, SIMBA \citep{Dave2019} uses stellar kinetic feedback with decoupled wind particles, HORIZON-AGN \citep{Dubois2016} uses thermal energy injection to model stellar feedback.
%%%%%%%%%%%%
%
In this paper, we use our own chemodynamical code \citep{Taylor2014} %hydrodynamical cosmological simulation TK2014
based on the \texttt{GADGET} hydrodynamical code \citep{Springel_yoshida_2001,Springel2005} to {systematically investigate} the effects of supernova feedback on the chemical evolution of galaxies. 
%We use an improved version of the code that contains several physical processes related to galaxy formation and evolution, such as radiative cooling, star formation, supernovae feedback \citep{Kobayashi_2007}, and black hole physics \citep{Taylor2014}. We implement four different feedback models: thermal feedback, thermal feedback with a kinetic velocity, stochastic feedback and mechanical feedback. The simulations are run for different parameters at the same resolution with the same initial conditions. We select a fiducial model parameter for each feedback method. We study and compare the effects on the star formation rate (SFR) and the mass-metallicity relation (MZR).

Measuring metallicity from observed spectra of galaxies is subject to many previous studies, is an ongoing effort, and is available for
%The main components considered are 
stellar populations \citep[e.g.,][]{Worthey1992,Conroy2013} and the ISM \citep[e.g.,][]{Maiolino_Mannucci2019,Kewley2008}. 
The stellar mass--metallicity relation (MZR) was first discovered in local elliptical galaxies by studying the colour-magnitude diagram \citep{McClure1968}. The relation for the ISM was first observed in a small sample of nearby star-forming galaxies by \cite{Lequeux1979}. Later on, using the Sloan Digital Sky Survey (SDSS), several authors derived a clearer MZR for stars (e.g., \citealt{Gallazzi2006,Zahid2017}) and the ISM \citep{Tremonti2004,Curti2020}, where galaxy metallicity increases with stellar mass.
%On the other hand, \cite{Tremonti2004} measured the gas phase MZR for $\sim$ 53000 star-forming galaxies in the local universe (z$\sim$0.1).
Various methods are used to infer the metallicity of the gaseous phase. The main ones are different calibrations with the photoionization models \citep{Kewley2008}, strong line calibration \citep{,Curti2020}, and direct method based on electron temperature \citep{Curti2023}.
%ck{[commented out below the paragraph on elemental abundance ratios as we don't discuss them in this paper]}
%Observational elemental abundance data is highly produced with ongoing and future surveys with multi-object spectrographs (MOS; e.g., APOGEE, HERMES-GALAH, Gaia-ESO, DESI, WEAVE, 4MOST, MOONS). Moreover, metallicity and some elemental abundance ratios within galaxies are measured with integral field unit (IFU) spectrographs (e.g., SAURON, SINFONI, CALIFA, SAMI, MaNGA, KMOS, MUSE, HECTOR, and NIRSpec from JWST).

In this paper, we implement and compare four models of stellar feedback in our cosmological simulations. The physical processes included in our simulation are described in section 2. The results of the different feedback models on the MZR are presented in section 3. Comparing with the observed MZRs, we discuss our results and give our perspectives in section 4.

%%%%%%%%%%%%%%%%%%%%%%%%%%%%%%%%%%%%%%%%%%%%%%%%
%%%%%%%%%%%%%%%%%%%%%%%%%%%%%%%%%%%%%%%%%%%%%%%%
\section{Model}
Our simulation code is based on the "GAlaxies with Dark matter and Gas intEracT 3" code known as \texttt{GADGET-3}  \citep{Springel2005}. It uses TreeSPH \citep{Hernquist_Katz1989}, which combines the smoothed particle hydrodynamics (SPH) \citep{Lucy1977,Gingold_Monaghan1977} to follow the gas dynamics, with the hierarchical tree algorithm to compute the N-body gravitational interactions. 
The long-range force is calculated with the PM-algorithm using Fourier techniques.
We use an improved version of the code that contains several physical processes related to galaxy formation and evolution, such as radiative cooling, star formation, supernovae feedback \citep{Kobayashi_2007}, and black hole (BH) physics \citep{Taylor2014}.\\
\subsection{Baryonic Physics}
Gravitational instability physics (i.e. dark matter structures) is an important starting point in galaxy formation models. However, one of today's greatest challenges in cosmological simulations is implementing the baryonic astrophysical processes that describe the galaxy population. The key difference between dark and baryonic matter is that the latter can dissipate energy through radiative processes. In what follows, we discuss the main processes involved in galaxy formation.\\ %(i.e. gas cooling, star formation, chemical enrichment and supernovae feedback).\\
\textbf{Radiative Cooling}:
Radiative cooling is a process that allows a space object to lose heat by thermal radiation. It uses the cooling function $\Lambda (T)$, which expresses gas cooling by thermal radiation. This function assumes that the gas is optically thin (i.e. an emitted photon can typically leave the cloud). An example of a set of cooling curves is given by \cite{Sutherland1993} where $\Lambda (T)$ has multiple peaks and valleys because the emission mechanisms are most efficient at specific temperatures. For instance, it is characterised by a big bump at low temperature produced by line radiation and a tail at high temperature (above $10^7$K) produced by Bremsstrahlung. We also include Compton heating. In this work, we use the same metallicity-dependent cooling function implemented by \citet{Kobayashi2004}, which is computed with the \texttt{MAPPINGS}-III software \citep{Sutherland1993}. \\
\textbf{Star Formation}:
In galaxy simulations the formation of star particles is only allowed in a gas that obeys certain conditions, stars are formed in cool dense gas. As in \citet{Kobayashi_2007},  we use the star formation criteria used in \citet{Katz1992}, which are: (1) Star formation is only allowed in convergent flows, (2) Star formation is only allowed in regions where the cooling time is less than the dynamical time (rapid cooling), and (3) The gas has to be locally Jeans unstable.\\ %{(sound crossing times)}.\\\\
\textbf{Stellar Feedback}:
It is observed that stars represent less than 10$\%$  of the baryonic matter in the observable Universe \citep{Madau_Dickinson2014}. However, according to the predictions of the cosmic microwave background (CMB) models, all the gas has already cooled and formed stars by today. This problem was recognised by the earliest models of galaxy formation (\citealt{White_Rees1978}, \citealt{Dekel_Silk1986}) where they suggested that this overcooling may be solved by the consideration of the supernova feedback. Supernova energy heats the gas, dispersing out of the galaxy and reducing the galaxy's baryon fraction, leading to a star formation inefficiency. %Today, several other processes, such as photoionisation, stellar winds, and AGN feedback, are recognised to reduce star formation. These processes remain unresolved in most cosmological simulations, and several studies follow their evolution at galaxy scales. \\
There are two classes of feedback mechanisms that retard star formation, ejective and preventive feedback. Ejective feedback ejects the gas from the ISM, while preventive feedback stops the gas from accreting into the ISM. 
On the contrary, dying stars and supernovae eject metals, which enhance cooling and star formation \citep[e.g.,][]{Kobayashi_2007}. We include these effects self-consitently. More detail is given in Sect. \ref{Sect_FB}.\\
\textbf{AGN Feedback}:
In addition to supernova feedback, feedback from AGNs is essential in suppressing star formation in massive galaxies \citep[e.g.,][]{Silk2013,Taylor2015}. In a star-forming galaxy (with cool and dense gas) where the BH is not yet active, part of the gas produces stars, and the other part falls into the BH. After accreting enough gas, the BH becomes active and ejects outflows and radio jets. This mechanism, known as AGN feedback, heats and pushes the gas away, which slows down star formation and stops the BH from growing. No more fuel will cause the BH to deactivate, meaning nothing will heat the gas anymore (for it to expand), so it returns to the initial BH stage. At this stage, the gas can cool again. However, stars only form if the gas is dense enough, which is usually not the case.

%\ck{[A few sentences on the black hole formation, see Taylor2014]}
As in \citet{Taylor2014}, AGN feedback in our simulation is modelled as (1) BH seed formation: the seed BHs are formed with the first stars, any gas particle with a density higher than the specified critical density and with zero metallicity ($Z=0$) is converted into a BH particle with a seed mass of $1000 h^{-1} M_\odot$. (2) Growth: The seed BHs grow by accreting gas and by merging with other BHs. %This accretion is not resolved in cosmological hydrodynamical simulations. 
(3) AGN feedback: In each timestep, a certain amount of energy is produced by the BH and is distributed in a thermal form to a fixed number of neighbour gas particles. 
\\
\textbf{Chemical Enrichment}:
In our simulations, a star particle is not a single star but a set of many. We consider a star particle as a simple stellar population (SSP, i.e. stars with the same age and metallicity but different masses) %Most hydrodynamical simulations
and include a chemical enrichment model that tracks the enrichment of the gas with all elements up to zinc. Oxygen, carbon and iron abundances 
are mainly produced by core-collapse SNe, AGB stars, and SNe Ia, respectively \citep{Kobayashi_Karakas_Lugaro2020}. 
The initial mass function (IMF) of stars is taken from \citet{Kroupa2008}.
%Most codes track multiple individual elements  \citep{Kobayashi2004}.  \textcolor{red}{A FAIRE}
We compute oxygen abundance for the ISM to compare with observations of metallicities weighted by SFRs. And we use total metals for stellar metallicities weighted by V-band luminosities. 

\subsection{Stellar feedback}\label{Sect_FB}
In hydrodynamical simulations, gas particles are affected by nearby star particles locally. We also follow the cooling of the gas particles after the feedback. These are fundamentally different from the loading factor in previous work  \cite[e.g.,][]{Belfiore2016, Lian_Thomas2018, Lin_Zu2023}, which is the measure for the average effect of or within the galaxy (see \citealt{Taylor2020} for comparison between simulations and observations).
In the following, we describe four feedback methods proposed for hydrodynamical simulations.
\\
\textbf{Thermal feedback}:
The classical stellar feedback method used in galaxy simulations consists of the distribution of thermal energy from supernova explosions into the surrounding gas (e.g., \citealt{Katz1992}), namely, to the neighbour particles in a fixed radius or
with a fixed number, using a smoothing kernel $W_j$.
 \begin{equation}
    \Delta E = E_{\rm SN} W_j
\end{equation}
Where $\Delta E$ is the weighted fraction of the supernova energy received by the  $j^{\rm th}$ gas particle, and $E_{\rm SN}$ is the total energy ejected by supernovae from an evolving star particle in a given time step.
In our simulations, $N_\mathrm{ngb}$ nearest neighbour gas particles are selected
and receive the supernova energy weighted by the smoothing kernel. Then at each time step,  the total ejected energy $E_{\rm SN}$ is divided accordingly to the weighting to heat the gas particles individually. 

Our simulations also include %thermal
hypernova feedback \citep{Kobayashi_2007}. Since the energy of hypernovae is more than ten times larger than the supernova energy ($10^{51}$ erg), the temperature increase can be much more significant and can reach $\sim10^6$ K. Once the gas particles are heated to this temperature, they do not cool rapidly due to the low cooling rate. As a result, this reduces the SFR significantly, and the hypernova feedback is expected to be more efficient than supernova-only feedback.\\
%%%%%%%%%%%%%%%%%%%%%%%%%%%%%%%%%%%
\textbf{Kinetic feedback}:
This method consists of implementing outflows where the energy input is partially converted to kinetic energy \citep{Navarro_White_1993}. The thermal energy ejected by each supernova explosion is partially reduced by a parameter $f$ ($0\le f \le1$), representing the fraction of energy distributed as kinetic energy. This model simulates a shocked gas with a kinetic kick of velocity $v$ such as:
\begin{equation}
    v=\sqrt{2f E_{\rm SN}\mathrm{W}_j/M_j}
\end{equation}
where $M_j$ is the mass of the $j^{\rm th}$ gas particle that receives the energy, and $E_{\rm SN}\mathrm{W}_j$ is the weighted fraction of the supernova energy received by the  $j^{\rm th}$ gas particle. This velocity is added to the original velocity of neighbour particles isotropically.\\
\textbf{Stochastic feedback}:
This approach was first implemented by \cite{kay_thomas2003} in galaxy simulations, and generalised by \cite{Dalla2012} to complete the thermal feedback method and efficiently suppress star formation.
Thermal feedback may be inefficient because the thermal energy is mostly radiated away before it can be turned into kinetic energy. This may be because the mass of the gas receiving the supernova energy is too large. 
Without hypernovae the energy emitted by supernovae is not enough to efficiently heat these gas particles; hence the gas temperature remains too low and the cooling time too short. Another reason for the thermal feedback inefficiency may be the lack of resolution: the energy is mainly distributed to a high-density gas because the simulation does not resolve the hot and low-density areas (which remain missing).

The temperature jump of the neighbour gas can be increased by reducing the mass of the heated gas with respect to the star particle. This can be done by reducing the number of heated gas particles or by specifying the temperature jump of the heated gas. The first idea may cause an issue if even one gas particle is too massive. The second way can be done using stochastic feedback, where the probability of the gas particle being heated depends on the star-to-gas mass ratio and the specified temperature jump.
For the stochastic feedback, the idea is to select a random number of nearby particles (instead of heating all the neighbour particles as in the thermal feedback model).

We use the same model as in \cite{Dalla2012} where we define an energy increase $\Delta e$ used to heat the neighbour gas particles.
\begin{equation}
    \Delta e = f \frac{E_\mathrm{SN}}{N_\mathrm{ngb}}
    \label{eq_De_sto}
\end{equation}
where $E_\mathrm{SN}$ is the total supernova energy ejected by an evolving star particle in a given timestep, $N_\mathrm{ngb}$ is the number of neighbour gas particles, and $f$ is a parameter which is introduced to hold the probability that each of the $N_\mathrm{ngb}$ particles receives an energy increase of $\Delta e$ (i.e., $\Delta e >  \frac{E_\mathrm{SN}}{N_\mathrm{ngb}}$, hence $f>1$).
$\Delta e$ is the total energy a single gas particle will receive from the supernovae independently from the distance to the star particle. To determine if a gas particle will receive the energy or not,  a random number $0<r<1$ is compared to the condition:\\
\begin{equation}
    r < \frac{E_\mathrm{SN}M_*}{\Delta e \sum_j^{N_\mathrm{ngb}} M_j}
    \label{eq_r_sto}
\end{equation}
with $M_*$ the mass of the star particle and $M_j$ the  mass of the $j^{\rm th}$ gas particle receiving the energy. The gas particle receives an energy increase of $\Delta_e$ only if this condition is satisfied.
Larger $f$ results in a smaller number of gas particles heated with a larger energy $\Delta e$. Note that $M_j$ and $M_*$ are not constant in our simulations.\\
%%%%%%%%%%%%%%%%%%%%%%%%%%%%%%%%%%%
\textbf{Mechanical feedback}:
Mechanical feedback takes into account the supernova shock wave applied to gas particles. This model uses assumptions depending on the structure of the ISM at small scales and its interaction with the supernova remnants. As the supernova shock wave propagates, it accelerates particles, radiating energy away. 
%The expansion is self-consistent \ck{self-similar??}, so the variables differ only in scale, meaning the \ck{good momentum amount [what do you mean?]} injected by the mechanical feedback scheme depends on the scale of the resolved SN remnant.
There are mainly three phases in the life of a supernova.: i) the free expansion phase,
ii) the adiabatic or Taylor-Sedov phase \citep{taylor1950,sedov1959}, where the expansion proceeds adiabatically into the surroundings, and radiative losses are negligible. And iii) the radiative or `snowplough' phase, where the gas temperature in the shock wave drops as the cooling function increases and the shock slows until it merges with the surroundings and disappears.

We implement mechanical feedback similar to \cite{Hopkins2014} and \cite{Smith2018}, where the supernova shock wave is considered to occur during the Sedov-Taylor phase of expansion, during which the shock wave is energy conserving. %(assuming that energy lost due to cooling is low).
As in the kinetic model, a fraction $f$ ($0\le f \le1$) of supernova energy $E_{\rm SN}$ is ejected in a kinetic form, but is converted to a momentum kick. The total momentum injected in the rest frame of the star particle is
\begin{equation}
    P_\mathrm{tot}=\sqrt{2m_{\rm ej}f E_\mathrm{SN}}
\end{equation}
with $m_{\rm ej}$ the total mass ejected by the supernovae in a given timestep. 
Following  \cite{Kimm2015}, the momentum as the remnant transitions to the snowplough phase is given as 
\begin{align}
    P_\mathrm{fin}&=  3\times 10^{10} \mathrm{km\:s}^{-1} \mathrm{M}_\odot \mathrm{\:} E_{51}^{16/17} n_\mathrm{H}^{-2/17} Z'^{-0.14}
\end{align}
with $E_{51}\equiv E_\mathrm{SN}/10^{51}$ erg, $n_\mathrm{H}$ is the hydrogen number density, and $Z'$ the metallicity in solar units $(Z'\equiv \max(Z/Z_\odot, 0.01))$.
The correct momentum, therefore, depends on the stage of the expansion. 
We calculate both forms of momentum in the code for each star particle during each timestep, and choose as
\begin{equation}
    \Delta P = \mathrm{W}_j P_\mathrm{tot}\, \min(\delta M, \delta P) \label{minDP} .
\end{equation}
Where $\delta M = \sqrt{1+\frac{m_j}{\Delta m_j}}$ is associated to the resolved Sedov-Taylor phase, with $m_j$ the initial mass of the $j^{\rm th}$ gas particle receiving the energy, and $\Delta m_j$ the mass received by the $j^{\rm th}$ gas particle from supernovae in nearby star particles. %the supernova shock wave}. 
And $ \delta P = \frac{P_\mathrm{fin}}{P_\mathrm{tot}}$ is associated with the unresolved exit of the Sedov-Taylor phase.
$\mathrm{W}_j P_\mathrm{tot}$ is the portion of momentum received by $j^{\rm th}$ gas particle.

%\ck{[Explain the meaning more. Smaller delM means what and what happens. Smaller delP means what and what happens.]}
%We calculate both forms of momentum in the code for each star particle during each timestep and apply the lowest of the two in eq. \ref{minDP}. 
%We compare the total and exit momentums to obtain the maximum momentum that can be input into the gas particle.
%\ck{Since $\Delta m_j<<m_j$, the smaller?? momentum between $P_\mathrm{tot}$ and $P_\mathrm{fin}$ is approximately used.}

\subsection{Initial conditions}

%\ck{IC is the same as in K07 but with updated cosmological parameters. Give mass and smoothing lengths.}

We use a $\Lambda$CDM cosmology with $h$ = 0.68, $\Omega_m$ = 0.31, $\Omega_\Lambda$ = 0.69 and $\Omega_b$ = 0.048 \citep{Plank2018}. The simulations presented in this paper are run at the same resolution with the same initial conditions (as in \citet{Kobayashi_2007} with updated cosmological parameters): Same number of dark matter and gas with a resolution of  $N_\mathrm{gas}$=$N_\mathrm{DM}$=$128^3$ with mass $M_{\rm DM}= 3.47 \times 10^{7} h^{-1} M_\odot$ and $M_{\rm gas}= 6.35 \times 10^{6} h^{-1} M_\odot$.
%\ck{[Read from snapshot000]}
The simulation is run in a periodic, comoving, cubic box volume of 10 $h^{-1}$Mpc on a side, with a gravitational softening lengths of $\epsilon_{\rm DM}$ = 1.6875 and $\epsilon_{\rm gas}$ = 0.84375 $h^{-1}$ kpc. We %generate star and gas particles and 
use the Friend-of-Friends (FoF) algorithm to locate galaxies as in \citet{Taylor2014}.

\subsection{Fiducial parameters}\label{fiducial}

We run our simulation with the same initial conditions and different parameter values $f$. For the kinetic feedback we run $f$=[0,1, 0.5, 1, 2, 3, 5, 10, 30, 50, 70, 90]$\%$ and find that the larger is $f$, the stronger the feedback. A large $f$ suppresses star formation too much, while $f<1$ gives very similar results as the thermal feedback (see Appendix \ref{SFR_fiducial_f}). For these reasons, we decide to use $f=1\%$ as our fiducial parameter. 
For the mechanical feedback, we run $f$=[1, 2, 5, 10, 30, 50, 70]$\%$, and with the same reasoning as for the kinetic feedback, we choose $f=1\%$ as our fiducial parameter.

For the stochastic feedback, $f$ is the fraction of total energy ejected from supernovae and is proportional to the probability that a gas particle is heated by the supernovae. A large $f$ is equivalent to a large $\Delta e$ which yields the right-hand side of Equation \ref{eq_r_sto} to be small. Therefore, for a large $f$, Equation \ref{eq_r_sto} is rarely satisfied, as only a small number of particles receive the energy increase and are impacted by the supernova feedback. We ran the values $f$ = [1, 3, 5, 10, 30, 50, 70, 90] (see Appendix. \ref{SFR_fiducial_f}) and found that for $f<50$ the impact of $f$ is not significant, and for $f>50$ the feedback is too weak. Therefore, we choose to use $f=50$ as our fiducial parameter for the stochastic feedback as it is our largest supernova energy fraction where enough particles are heated for the feedback to be effective.

%%%%%%%%%%%%%%%%%%%%%%%%%%%%%%%%%%%%%%%%%%%%%%%%%%%%%%%%%%%%%%%%%%%%%%
%\input{section3/results}
\section{results}
\subsection{Density and temperature evolution}

Figure \ref{rho_maps} shows the redshift evolution of the gas density in our cosmological simulations from the same initial conditions for the four feedback models with our fiducial parameters. At $z=2$ (bottom row), the density distribution is similar for all models. 
At $z=1$ (middle row), the kinetic feedback starts to behave differently. For example, if we focus on the top left region of the map, we can notice a large ring-like structure that is not occurring for the other models.
At $z=0$ (top row), one can distinguish a rich filamentary structure for the thermal, stochastic, and mechanical feedbacks (columns 1, 2, and 4, respectively); however, with the kinetic feedback, the density becomes very diffuse. There is no significant difference in the dark matter structure.
The gas in our simulations is accreted along the filaments falling toward a central node with higher density; this triggers star formation, enhances supernova feedback, and drives galactic winds. The supernova feedback starts earlier in the kinetic model, which explains the ``rings'' at $z=1$. This pushed gas keeps moving away from galaxies, which causes the diffuse structure at $z=0$.
There are no significant differences among the three other models.

Figure \ref{map_T}  shows the gas temperature in our simulations for the four models at $z=0$. Here again, we see a drastic behaviour with the kinetic feedback, which overheats the gas above $10^6$ K. The temperature is similarly high also with the stochastic feedback. Therefore, it is expected 
that the stochastic feedback model will have diffuse density structure 
similarly to the kinetic feedback model in the future time. 
On the other hand, the gas temperature in the thermal and mechanical feedback models ranges from $\sim$2000K in low-density regions to $\sim$ 6000K in dense regions. Overall, the mechanical feedback results in colder gas, and the cold areas are more extended than in the case with the thermal feedback.

%\ck{Figure 3??}

Figure \ref{map_Z}  shows the gas metallicity for the four models at $z=0$. 
The metallicity is distributed similarly for the thermal, stochastic and mechanical feedbacks, but stochastic feedback gives slightly less extended metallicity distribution.
The supernova feedback enhances the production of galactic winds, which mainly enriches the ISM and only slightly the IGM. 
On the other hand, the kinetic feedback produces much stronger galactic winds and strongly enriches the IGM.

\begin{figure*}
	\includegraphics[width=\textwidth]{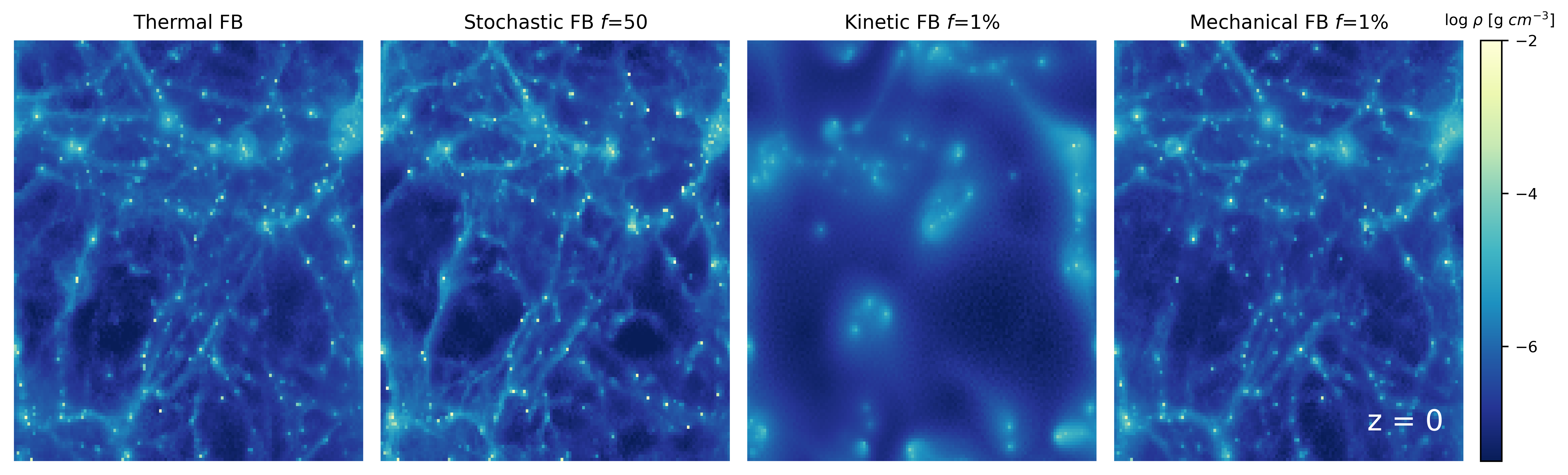}
\end{figure*}
\begin{figure*}
	\includegraphics[width=\textwidth]{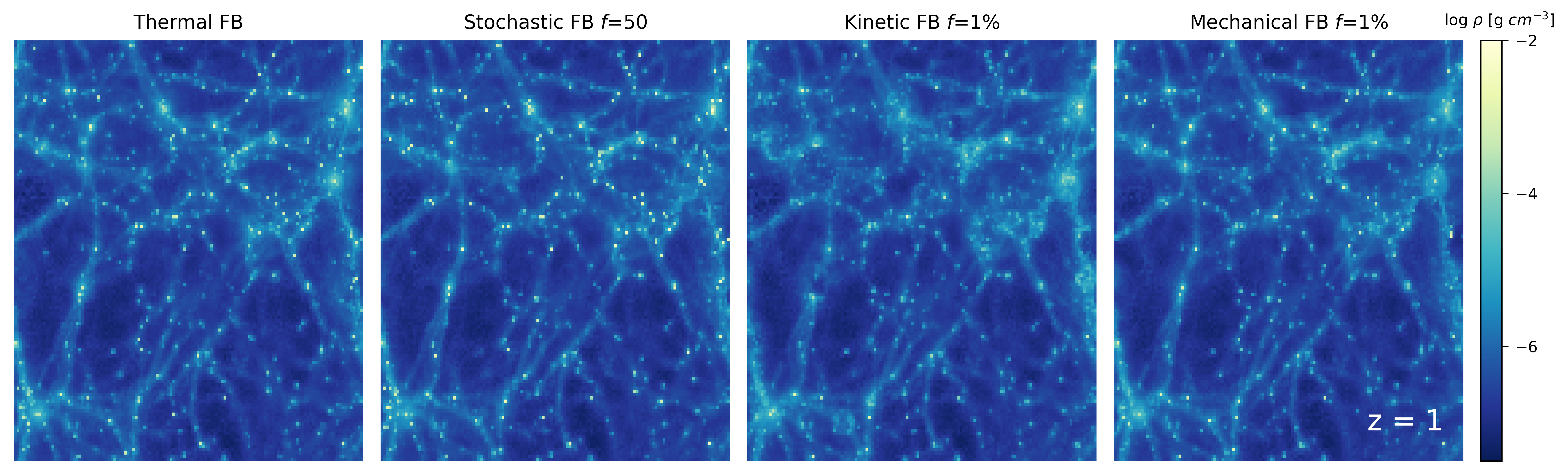}
\end{figure*}
\begin{figure*}
	\includegraphics[width=\textwidth]{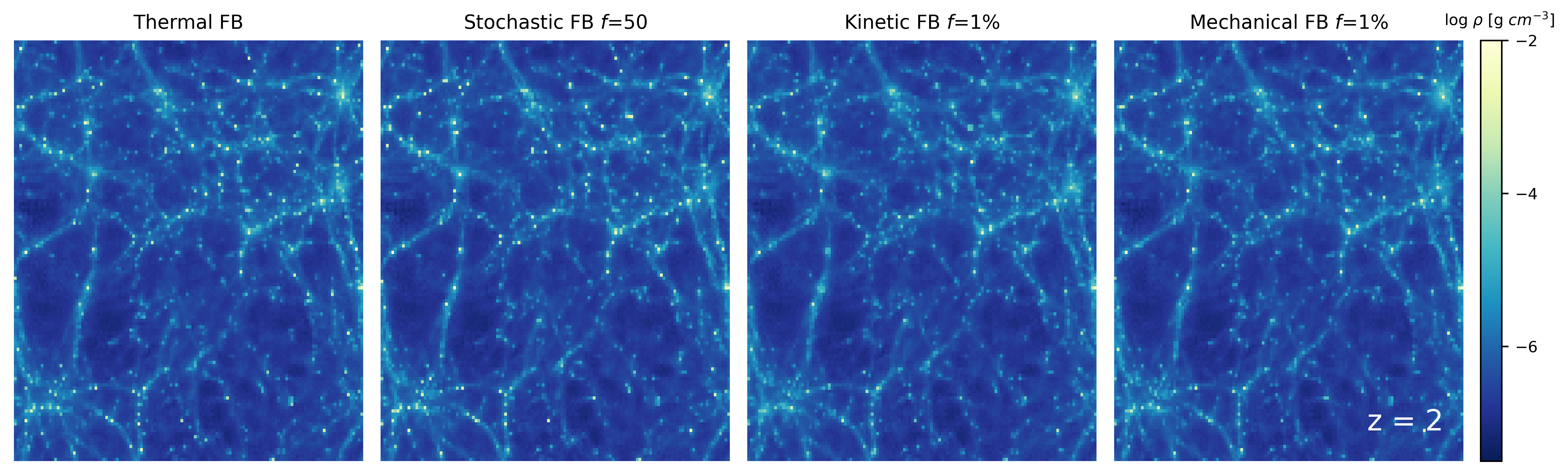}
    \caption{Density evolution of our cosmological simulations in 10 $h^{-1}$ Mpc$^3$ box for our four feedback (FB) models: thermal, kinetic, stochastic, and mechanical in the 1st, 2nd, 3rd, and 4th columns, respectively, with the fiducial parameters in section \ref{fiducial}. We show projected gas density at $z=0, z=1$, and $z=2$ in the top, middle and bottom rows, respectively.}\label{rho_maps}
\end{figure*}
%%%%%%%%%%%%
\begin{figure*}
	\includegraphics[width=\textwidth]{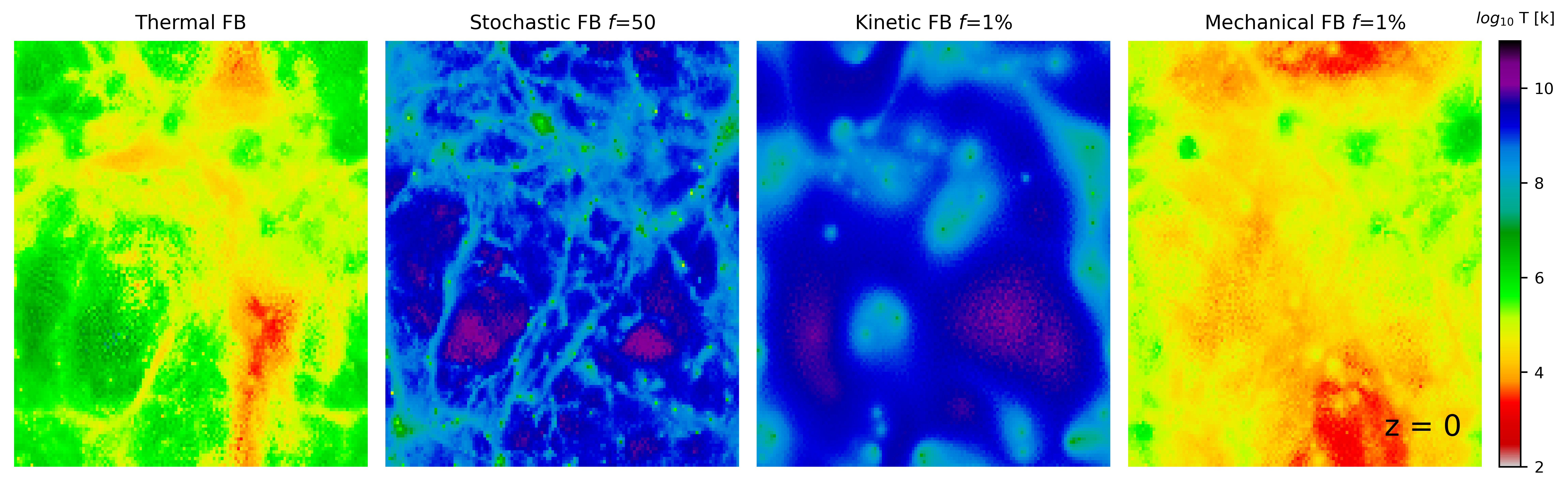}
    \caption{The same as Figure \ref{rho_maps} but for temperature maps of our cosmological simulations for the four feedback models at $z=0$.}\label{map_T}
\end{figure*}
\begin{figure*}
	\includegraphics[width=\textwidth]{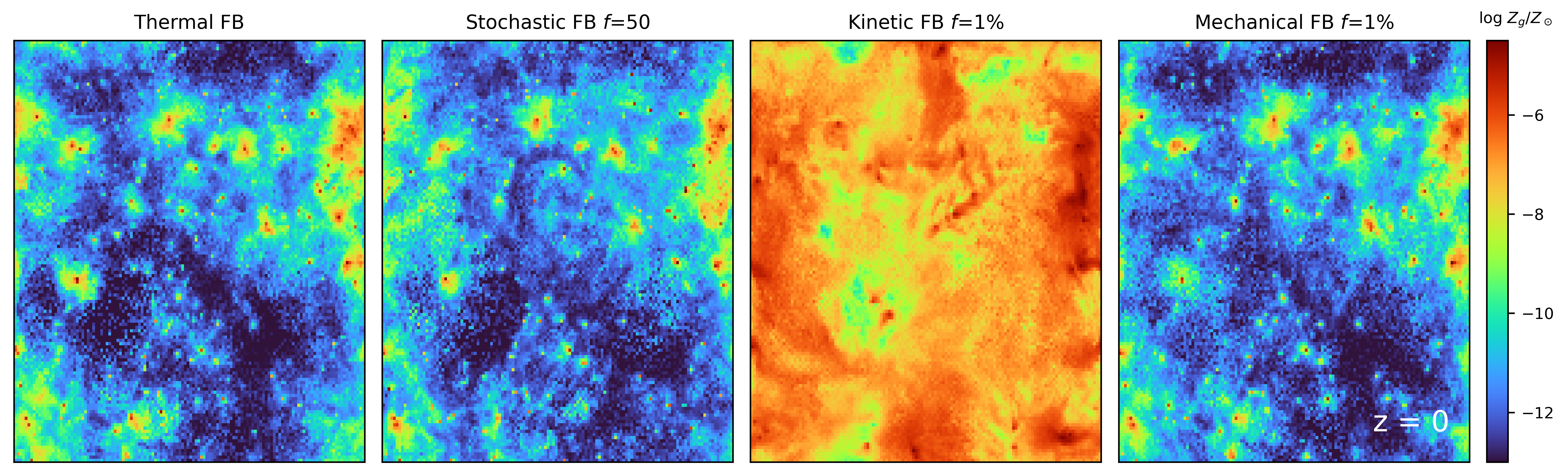}
    \caption{The same as Figure \ref{rho_maps} but for the gas-phase metallicity, $\log Z_{\rm g}/Z_\odot$, in our cosmological simulations for the four feedback models at $z=0$.}\label{map_Z}
\end{figure*}
%%%%%%%%%%%%%%%%%%%%%%%%%%%%%%%%%%%%%%%%%%%%%%%%%%%%%%%%%%%%%%%%%%%%%%
%%%%%%%%%%%%%%%%%%%%%%%%%%%%%%%%%%%%%%%%%%%%%%%%%%%%%%%%%%%%%%%%%%%%%%
\newpage
\subsection{Gas-phase diagram}\label{sect_diag}
\begin{figure*}
	\includegraphics[width=\textwidth]{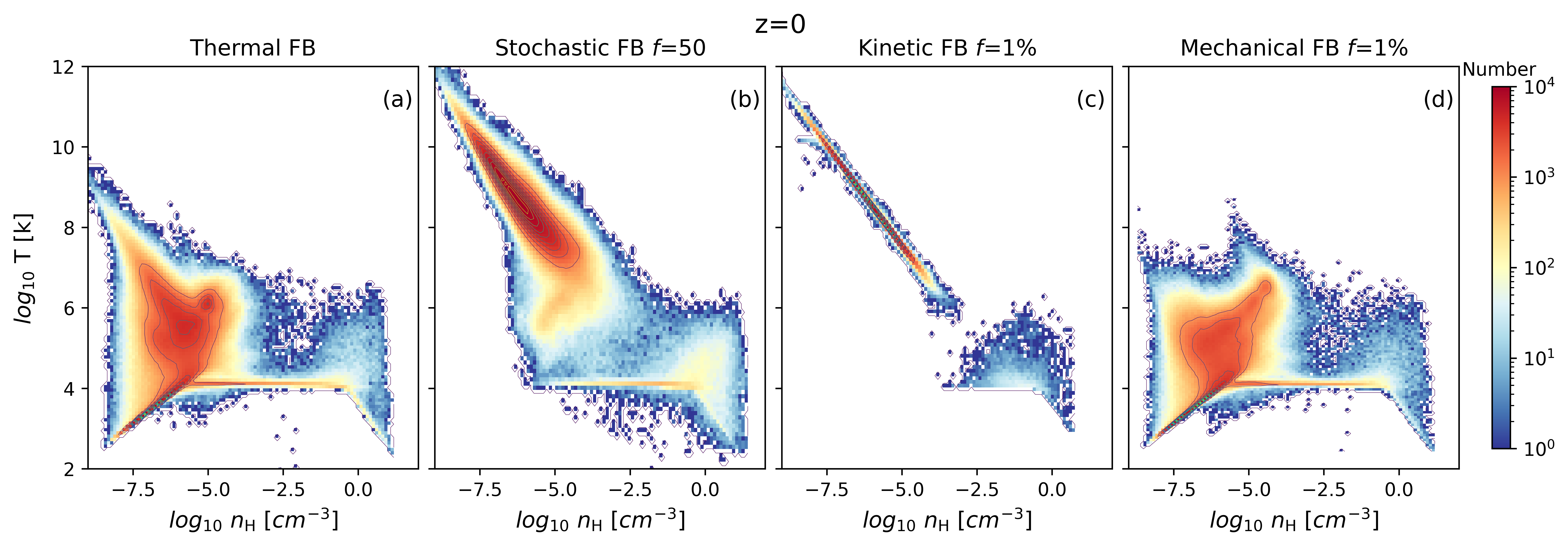}
    \caption{Density-temperature phase space diagrams for thermal (panel a), kinetic (panel b), stochastic (panel c), and mechanical (panel d) feedback models. Each panel shows the temperature as a function of hydrogen number density with the colour contour indicating the number density of the gas particles in the entire simulation volume at $z=0$.
    %\ck{[Unit correct?]}
    }\label{gas_pha_diag}

	\includegraphics[width=\textwidth]{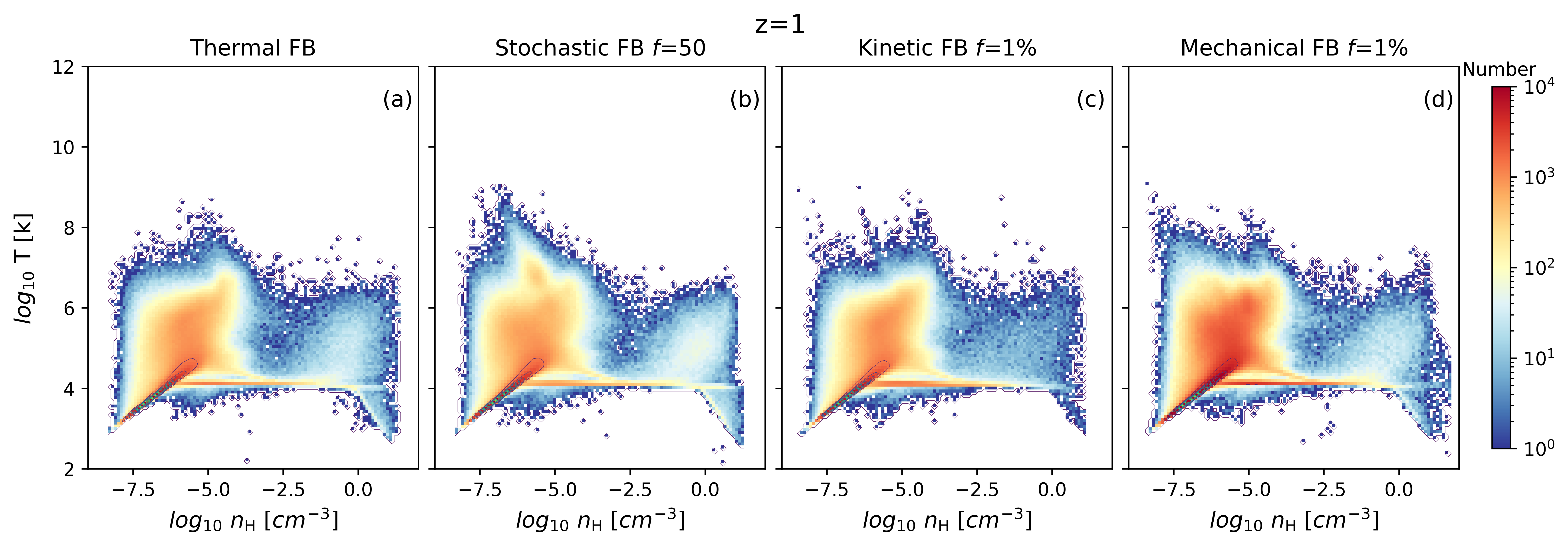}
    \caption{The same as Figure \ref{gas_pha_diag} for $z=1$.}\label{gas_pha_diag_z1}
\end{figure*}

As described in \cite{Dave2001}, the baryons in the universe are found in four different regions of the gas phase diagram: (1) The \textit{diffuse region} (low temperature $T<10^5$K and low density $\rho$) contains adiabatic gas outside galaxies with no specific role.  
(2) The \textit{condensed region} (low temperature $T<10^5$K and high density $\rho$) contains stars and cool gas inside galaxies.
(3) The \textit{hot region} ($T>10^7$K) contains the hot gas in galaxy clusters. 
(4) The \textit{warm-hot} region ($10^5K<T<10^7$K) contains the baryons in the IGM.
%\ck{Such matter around the galaxy can be observed with metal absorption lines can called circum-galactic medium (CGM, Peroux ARAA).}
Such matter surrounding galaxies more closely can be observed with metal absorption lines and called CGM \citep{perou2020ARAA}.

Figure \ref{gas_pha_diag} shows the density-temperature phase space diagram at $z=0$ for the four feedback models with our fiducial parameters. 
The main noticeable feature is the different behaviours of the warm hot region. The diagrams of the thermal and mechanical models show two bumps at two different temperatures: the first bump is at $T\sim10^4$K which corresponds to the peak of hydrogen cooling temperature, and the second bump is at $T\sim10^5$K which corresponds to the peak of helium cooling.

The straight horizontal line at temperature $T\sim10^4$K is caused by the cooling function sharply dropping below $T\sim10^4$K (see Fig.13 of \citealt{kobayashi_taylor_2023}), which prevents gas cooling until it reaches very high density.
Note that molecular cooling is not included, which would weaken this behaviour.
%not physical but systematic. It is due to the cooling function used in the simulation, which becomes zero at $T<10^4$K because it does not include molecular cooling except for the metal-rich case, which can cool down to lower temperatures. 
%In the code, the minimum temperature is set to be $T=100K$. In a metal-rich case, our simulation allows the gas particles to cool down to $T\sim 100K$ because metal-enhanced gas is considered to have dust cooling and molecular formation below $10^4K$.
At high densities, there are more star-forming particles in the stochastic feedback model than in the other models. % -- pls check.

The gas phase temperature-density diagram for the kinetic feedback has a strange behaviour as most of the gas is hot and diffuse; this is due to the cooling function used in our simulations, which drops down when the particles are heated beyond $\sim 3\times10^5$ K. %(i.e., when the particles reach this temperature they will not cool down). 
This figure supports the prediction that the stochastic feedback will end up with the same behaviour as the kinetic one. 
These diagrams indicate that the mechanical feedback is a better model for this resolution. 

%\ck{Comments on Fig5?}\\
The drastic changes for the kinetic and stochastic feedback happen after $z=1$ (see section \ref{redshift_evolution}), and Fig. \ref{gas_pha_diag_z1} shows the gas phase diagram of the four models at $z=1$ where there is no significant difference between the models. If we look closer, the mechanical feedback has a slightly larger amount of warm gas (at T$\sim 10^5$ K); this is due to its efficiency, making the non-star forming gas particles either heated or ejected. 

\newpage
\subsection{Cosmic Star Formation Rate}
\begin{figure}
	\includegraphics[width=\columnwidth]{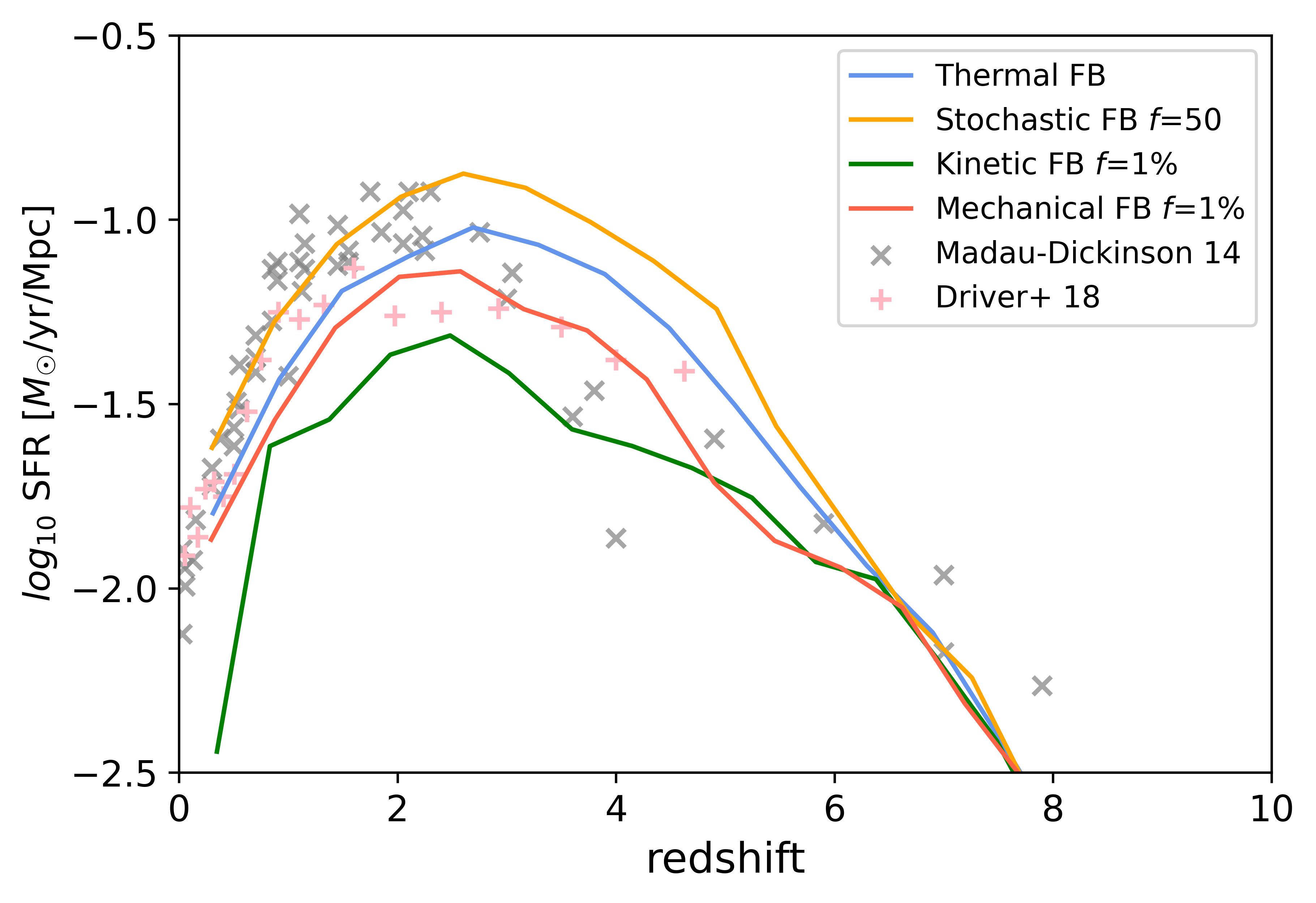}
    \caption{Cosmic star formation rate history of our 10 $h^{-1}$ Mpc simulations with our different feedback models: thermal (blue), stochastic (orange), kinetic (green), and mechanical (red). The observational data are taken from \protect\cite{Madau_Dickinson2014} (grey cross) and \protect\cite{Driver2018} (pink plus).}
    \label{cosmic_SFR}
\end{figure}

Figure \ref{cosmic_SFR} shows the cosmic SFR history obtained with each of the four feedback models with fiducial parameters. The SFR increases with time until the cosmic noon at redshift $\sim2$, where the SFR was at its maximum. It decreases from $z \sim$ 2 to the present day because most of the cold gas has already turned into stars,  but also because the formation of stars is suppressed by the presence of more supernova and AGN feedback. These test runs with a limited box size do not have very massive galaxies and galaxy clusters, which may explain our SFRs are lower at $z \ltsim 2$ than observed. 
The stochastic feedback shows a similar behaviour with a slightly higher SFR (due to weaker feedback). In the kinetic case, the feedback impact can only be seen after sufficient star formation has occurred (i.e. at $z \ltsim$ 6). After $z=6$, the feedback is too strong, and star formation is suppressed too much, compared with the observations.
We retrieve the same behaviour for the mechanical feedback with a less strong suppression of SFRs.
Observational data are taken from \cite{Madau_Dickinson2014} (grey cross) and \cite{Driver2018} (pink plus).
%\ck{[At the JWST conference, I saw a paper that showed a peark at higher redshift. Do you remember? It would be good to plot that results too.]}

\newpage
\subsection{Redshift evolution}\label{redshift_evolution}
\begin{figure*}
	\includegraphics[width=\textwidth]{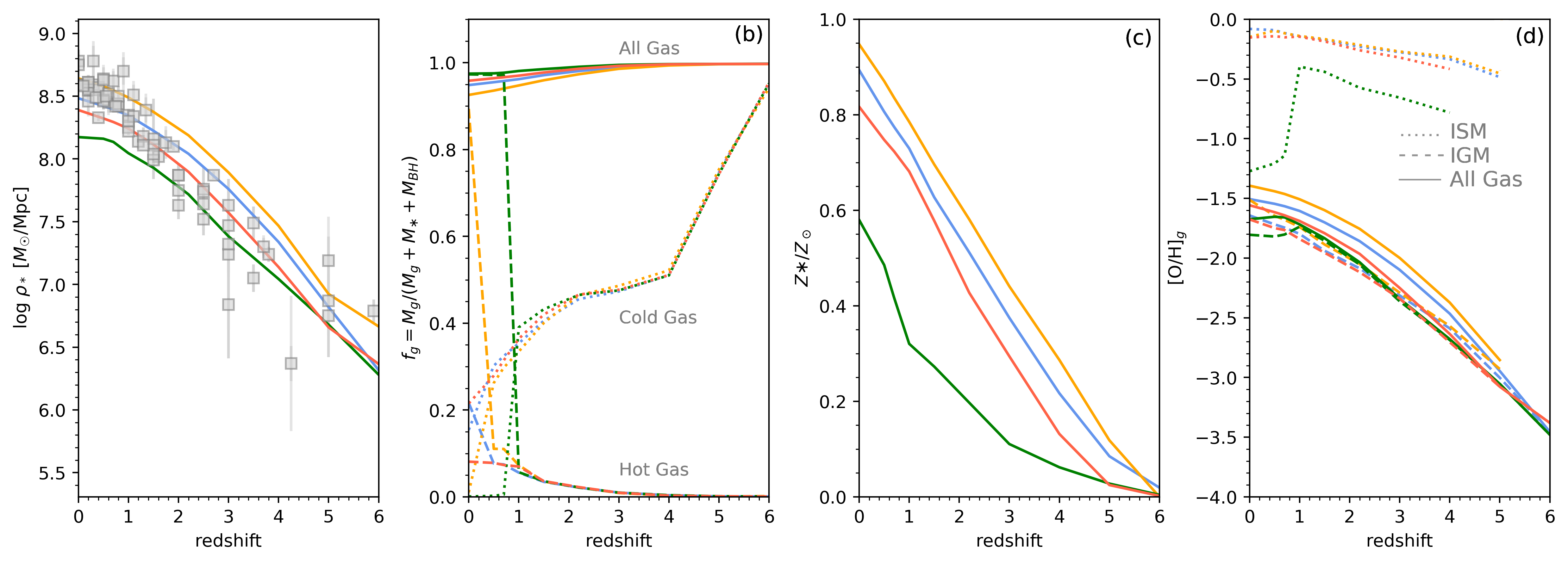}
    \caption{(a) Stellar mass density as a function of redshift, comparing to the observational data (grey square) taken from \protect\cite{Madau_Dickinson2014}. (b) Cosmic gas fraction $f_{\rm g} \equiv M_{\rm g}/(M_{\rm g}+M_*+M_{\rm BH})$ for all gas (solid lines), hot gas ($T > 10^6$K, dashed lines), and cold gas ($T < 1.5 \times 10^4$K, dotted lines). 
    %\ck{[Can you analize z=0-1.5 with finer grid, so that we can see where the change happens?]}
    (c) Cosmic stellar metallicity evolution. (d) Gas-phase oxygen abundances evolution for all gas (solid lines), ISM (dotted lines), and IGM (dashed lines). In all panels, the thermal, stochastic, kinetic, and mechanical feedbacks are always shown in blue, orange, green, and red, respectively. 
    %\ck{[Put (a) etc inside the figures. Delete ``(HR)'']}
    }\label{z_evolution}
\end{figure*}

%%%%%%%%%%%%%%%%%%%%%%
In Figure \ref{z_evolution}, we investigate the redshift evolution of different cosmic quantities of the gas and stars in our simulations.
The panel \ref{z_evolution}a shows the stellar mass density as a function of redshift, obtained with our four feedback models. The kinetic model has the lowest stellar density at all redshifts plotted here since, as shown in the cosmic SFR, it is the strongest feedback that suppresses star formation the most. Our simulations follow a similar trend as observational data from \cite{Madau_Dickinson2014}. However, at $z>1$, the kinetic and mechanical feedback seems to fit better,  while at $z<1$, the thermal and stochastic feedback work better. 

The panel \ref{z_evolution}b shows the gas fraction defined as $f_{\rm g}$ $\equiv$ $M_{\rm g}/(M_{\rm g} + M_* + M_{\rm BH}$) as a function of redshift for the four feedback models. At $z>4$, the total gas fraction (solid lines) is $f_{\rm g} \sim 100\%$ for all models. From $z=4$,  the total gas fraction decreases to $f_{\rm g}=94\%$, $92\%$, $97\%$, and $96\%$ at $z=0$, respectively for thermal, stochastic, kinetic, and mechanical feedback. All these are comparable with the observational estimates (e.g., $f_{\rm g}$ = 0.91-0.95 in \citealt{Madau_Dickinson2014}).
%\citealt{Fukugita_Peebles_2004} where $f_{\rm g}$ = 0.95 $\pm$ 0.09) \ck{[Check Peroux's ARAA for more recent result]}.
The stochastic feedback gives the smallest total gas fraction, and despite the large amount of hot gas in Figure \ref{gas_pha_diag}, %probably because it turned 
more gas is turned into stars overall (as shown in Fig. \ref{z_evolution}a).
The kinetic feedback has the largest total gas fraction.

The dashed and dotted lines show the hot and cold gas fractions with $T>10^6$ and $T<1.5\times10^4$ K, respectively.
At $z>6$, most of the gas was cold in all models. The cold gas fraction decreases with time while the hot gas fraction increases, mainly due to stellar and AGN feedback. At $z\le$2, the cold gas fraction differs depending on the feedback models. The kinetic model has the highest cold gas fraction, which may be explained by the lower SFR than the other models. The gas is drastically heated in the kinetic model at exactly $z=1$ due to the high efficiency of stellar feedback, \di{this transition is not caused by our redshift binning but is real due to the nature of this feedback model. Around this redshift, a large number of gas particles are heated above $10^5$ K, where the cooling rate is low (as explained in section \ref{sect_diag}), and are ejected from galaxies, suddenly increasing the hot gas fraction.} A similar behaviour is observed for the stochastic feedback later on at $z=0.5$, however, it is less sharp and evolves up to $z=0$. This extreme temperature change starts exactly at $z<1$ and $z<0.5$ for the kinetic and stochastic feedback, respectively. And the difference between $z=0$ and $z=1$ is clearly observed in the gas-phase density-temperature diagrams in Fig. \ref{gas_pha_diag} and Fig. \ref{gas_pha_diag_z1}.  
%\ck{[Might be useful to show the figures at $z=1$ for this sentence.]}

The panel \ref{z_evolution}c shows the stellar metallicity across cosmic time for the four models, which increases as time followed by star formation. The kinetic model has the lowest stellar metallicity at all redshifts since fewer stars are produced than in any other models. It is clear that the supernova feedback model has an impact on the present-day stellar metallicity, with $Z_*/Z_\odot$= 0.89, 0.94, 0.57, 0.81 at $z=0$ for thermal, stochastic, kinetic, and mechanical feedback, respectively.

The panel \ref{z_evolution}d shows the evolution of oxygen abundance with time with the four feedback models for all the gas (solid lines), the ISM (dotted lines) and the IGM (dashed lines), separately. As defined in \cite{Kobayashi_2007} and \cite{Taylor2016}, the ISM is all gas particles in galaxies identified by the Friend of Friend algorithm \citep{Springel2001}, and the IGM is all the other gas particles. The kinetic feedback has a lower oxygen abundance because it has less star formation, therefor fewer heavy elements are produced by supernovae. From $z=1$, the oxygen abundance is reduced in the ISM due to the winds that eject the oxygen-enhanced gas outside the galaxy, but also due to dilution, where all matter is mixed up and fills the ISM with hydrogen, which explains the drop in the plot. 
Table \ref{table_z_evolution} summarizes the values of the cosmic stellar mass density $\log \rho_\ast$, gas fraction $f_{\rm g}$ (for all gas, hot gas, and cold gas), stellar metallicity $Z_\ast /Z_\odot$, and gas-phase oxygen abundance $[\mathrm{O/H}]_{\rm g}$ (for all gas, ISM, and IGM) at $z=0$, for the four feedback models.

%\ck{[Make a table of z=0 numbers of all four models inclusing gas metallciities.]}

\begin{table}
\centering
\begin{tabular}{ |l|c c c r| } 
\hline
FB Model & Thermal & Stochastic & Kinetic & Mechanical \\
\hline
 \hline
 $\log \rho_\ast$   & 8.483     & 8.642      & 8.173     & 8.389       \\
% \hline
\hline
$f_\mathrm{g}$ (All Gas)      & 0.948     & 0.925      & 0.974     & 0.958       \\
% \hline
$f_\mathrm{g}$ (Cold Gas)  &  0.154    &  0.0103     &  0.0007    & 0.215    \\
% \hline
$f_\mathrm{g}$ (Hot Gas)  &  0.216    &  0.893     &    0.973  &  0.081   \\
%\hline
\hline
$Z_\ast /Z_\odot$ &    0.893  &    0.947   &   0.974   &   0.816  \\
 %\hline
 \hline
$[\mathrm{O/H}]_\mathrm{g}$ (ISM)   & -0.080     & -0.148      & -1.272     & -0.152   \\
% \hline
$[\mathrm{O/H}]_\mathrm{g}$ (IGM)  & -1.643     & -1.512      & -1.805      & -1.676      \\
% \hline
$[\mathrm{O/H}]_\mathrm{g}$ (All Gas) & -1.506     & -1.393      & -1.674     & -1.557       \\
 \bottomrule
\end{tabular}

\caption{Cosmic stellar mass density $\log \rho_\ast$, gas fraction $f_{\rm g}$ (for all gas, hot gas, and cold gas), stellar metallicity $Z_\ast /Z_\odot$, and gas-phase oxygen abundances $[\mathrm{O/H}]_{\rm g}$ (for all gas, ISM, and IGM) of the thermal, stochastic ($f=50$), kinetic ($f=1\%$) and mechanical ($f=1\%$) feedback models at $z=0$.}\label{table_z_evolution}
\end{table}

\subsection{Mass--Metallicity Relations}
\subsubsection{Stellar Populations}
\begin{figure}
\centering
	\includegraphics[height=9.5cm]{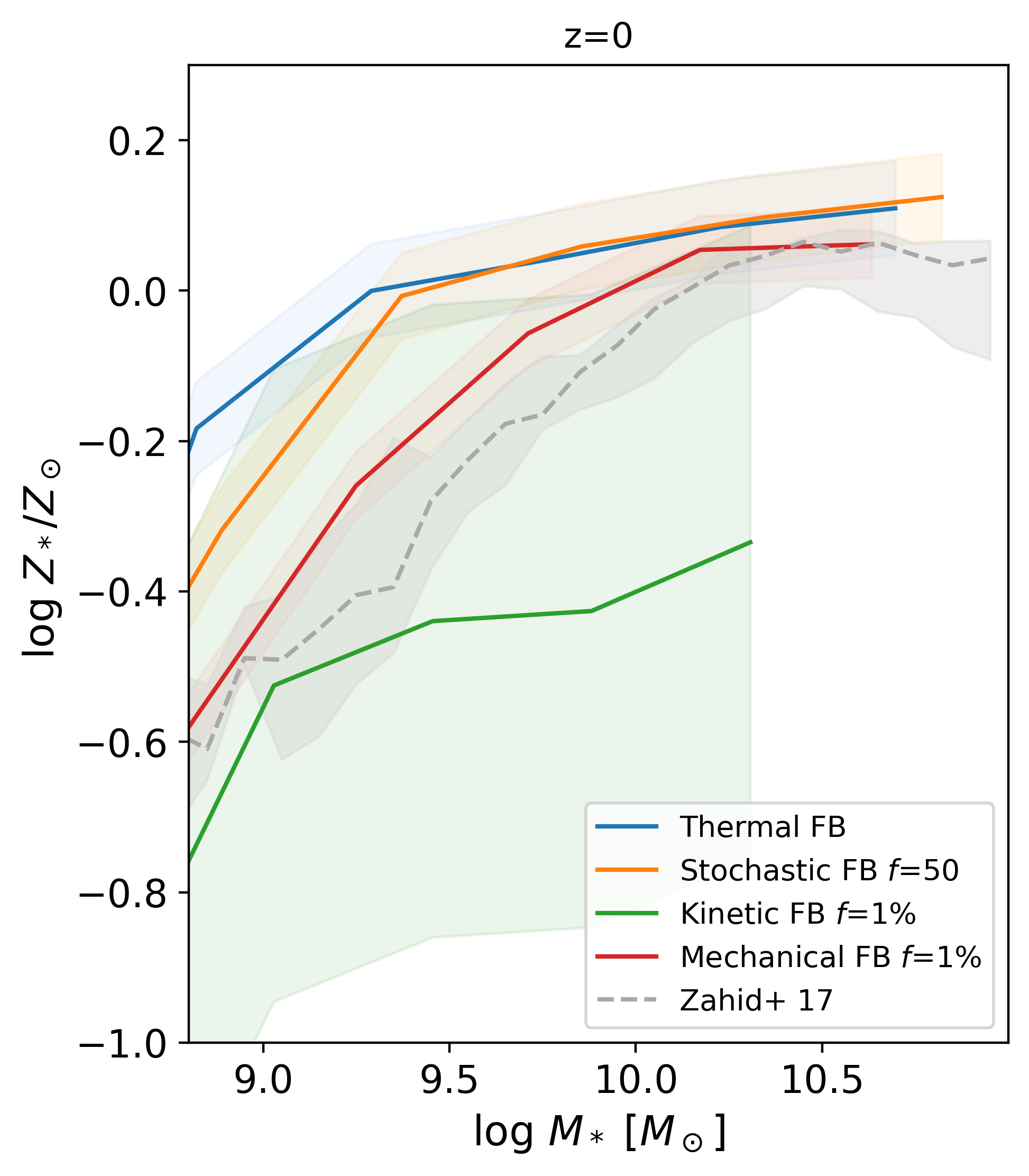}
    \caption{Stellar mass--metallicity relations with thermal (blue), stochastic (orange), kinetic (green), and mechanical (red) feedback models.  The stellar metallicity is V-band luminosity-weighted. The lines are for medians and the shaded areas show the 1$\sigma$ scatter.
    The observational data are taken from \protect\cite{Zahid2017} (grey dashed line, with 1$\sigma$ scatter). 
    %\ck{[Double-check - is this median, not average?]}
    }
    \label{stellar_MZR}
\end{figure}

Figure \ref{stellar_MZR} shows the stellar mass--metallicity relations (MZRs) for the four feedback models, with the integrated metallicity of stars in galaxies weighted by the V-band luminosity of star particles. 
In our simulations, a star particle is not a single star but a set of many. We consider a star particle a simple stellar population (SSP, i.e. stars with the same age and metallicity but different masses). 
V-band luminosities of star particles are calculated using the Binary Population and Spectral Synthesis (BPASS) code version 2.2.1 \citep{Stanway2018}.
The stellar metallicity of galaxies is measured in a 15 kpc projection from the galactic centre. The lines in Figure \ref{stellar_MZR} represent the median of the simulated galaxies, while the shaded areas display the $1\sigma$ scatter.
%\ck{[Also, explain how the lines are calculated]} 
 The solar metallicity used in the figure is $Z_\odot = 0.015$. 

%\ck{[Did you explain the SSP model you used? If not, a few sentences here. Also, explain how the lines are calculated. What is the solar metallicity value you used? Also, is it possible to show the $1\sigma$ scatter with shaded areas?]} 

Our thermal and stochastic models tend to overproduce metals compared with the local observations (black dashed line, with the grey shade for 1$\sigma$) taken from \cite{Zahid2017}. %\ck{although there is an uncertainty in the observations (refs)}.
Our kinetic feedback is not producing enough metals due to the lower SFR and not keeping enough metals in stars because of the kick velocity that drives the metals out of the galaxy. 
Among our four models,
mechanical feedback gives the closest matches to the observed relation
%is the model that matches the most with observations
from \cite{Zahid2017} at $z=0$, with this resolution.
%However, 
We aim to confirm this by running even higher resolution in a larger volume of cosmological simulations in our future work.

\subsubsection{Gas phase}
\begin{figure}
\centering
	\includegraphics[height=9.5cm]{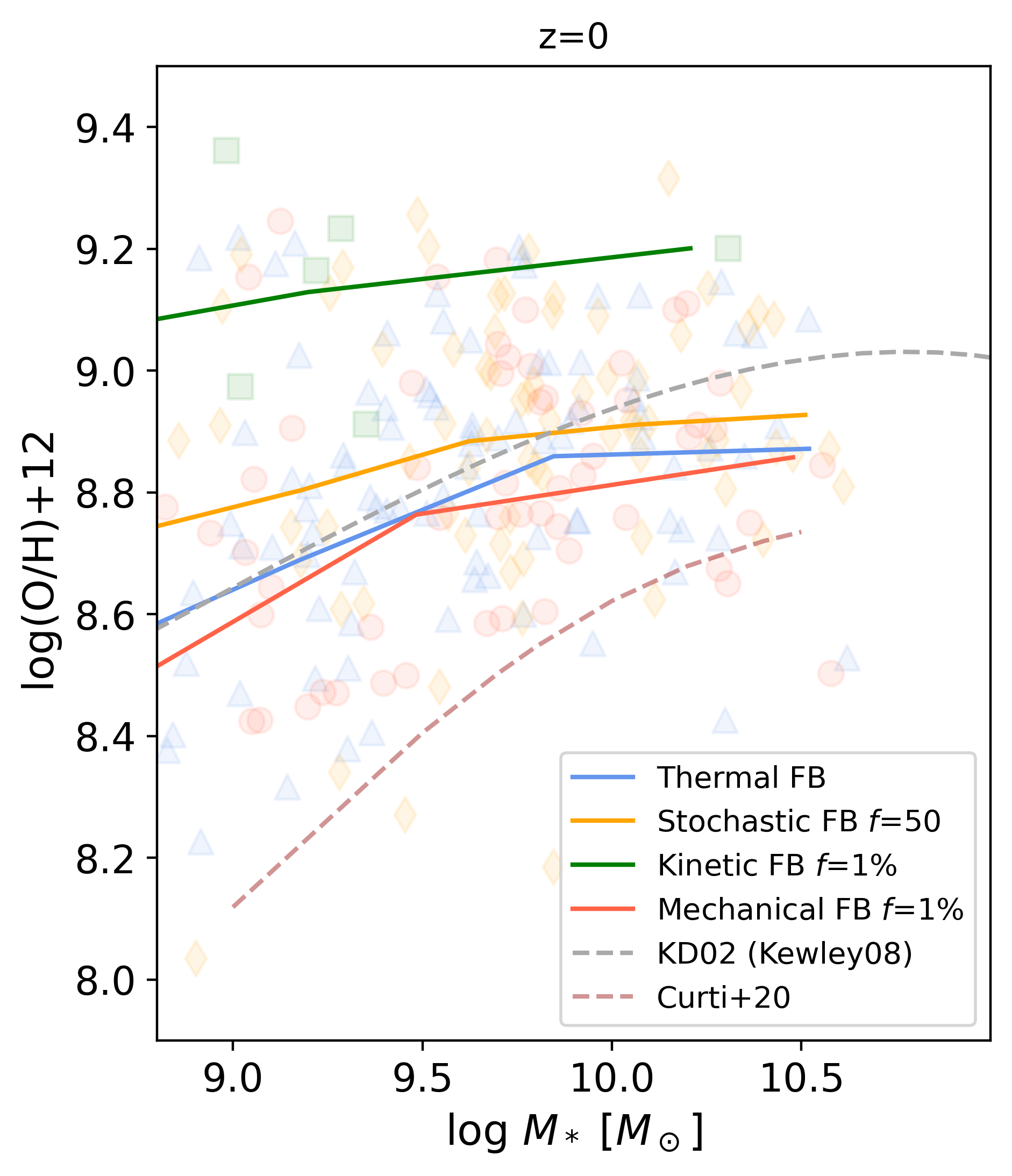}
    \caption{Gas-phase mass--metallicity relations with thermal (blue; triangles), stochastic (orange; diamonds), kinetic (green; squares), and mechanical (red; circles) feedback models. The SFR-weighted, gas-phase oxygen abundances of galaxies (number ratios relative to hydrogen) are shown. The observational data (grey dashed line) are from \citet{Tremonti2004} with the `KD02' scale in  \citet{Kewley2008}, \di{and from \citet{Curti2020} (brown dashed line).}
    }
    \label{gas_mzr}
\end{figure}

Figure \ref{gas_mzr} shows the gas-phase MZRs with our four feedback models. We calculate the gas-phase ``metallicity'' of galaxies by measuring the gas oxygen abundance within 15 kpc from each galactic centre, weighted by the SFRs of gas particles to compare with observations, which are weighted by emission lines. 
Not many gas particles are forming stars with the current simulation volume and resolution, particularly at the massive end. Therefore, we have limited data points for SFR-weighted gas-phase metallicities. 
Consequently, we show the metallicities of galaxies (points), in addition to a fit (linear fit of medians; solid lines) in Fig. \ref{gas_mzr}. The solar oxygen abundance adopted for our nucleosynthesis yields is $\sim$ 8.76.
%From this plot, we can conclude that for relatively high-mass galaxies at $\sim 10^{10}M_\odot$, the kinetic feedback has the highest gas-phase metallicity, which is unexpected and is possibly due to the {strong} ejection of metal-poor gas. %due to the metal outflow with the kick velocity. 
%There is no MZR for the gas-phase with the kinetic feedback at z=0, probably due to the strong impact on the stellar distribution, removing star particles from the galaxies with gas metallicity values.
\di{ At $z=0$, the kinetic feedback gives the highest gas-phase metallicity ($\sim$  9.1 dex), which is possibly due to the strong ejection of metal-poor gas. %This model only produces low-mass star forming galaxies (up to $\sim 10^{9.5}M_\odot$) due to the strong feedback that suppresses star formation, which explains the absence of kinetic MZR for massive galaxies in Fig. \ref{gas_mzr}.
From this figure (namely scatter plot), we can conclude that for relatively low-mass galaxies at $\sim 10^{9}M_\odot$, the thermal and mechanical feedbacks are in reasonably good agreement with the observed gas-phase MZR from \cite{Kewley2008} (grey dashed line). %at the plotted mass range.
The stochastic feedback seems to give a shallower slope than observed.
With the mechanical feedback, higher-mass galaxies ($\sim 10^{10}M_\odot$) tend to have slightly lower metallicities than in \cite{Kewley2008}, and are more comparable with \cite{Curti2020}'s observation (brown dashed line).}

%The stochastic feedback tends to give higher metallicities than observed probably with the same reason. 

%As for stellar MZRs, the scatter plot of the mechanical feedback \di{(red circles)} is in reasonably good agreement with the observed gas-phase MZR (grey dashed line) from \cite{Kewley2008}, at the plotted {mass} range. %compared with observational data .
%\ck{[Can you also shoe the redshift evolution of the MZRs?]}

It is important to note that while the shape of the stellar and gas-phase MZR is relatively robust against the methods used for the metallicity determination, the absolute amplitude of the stellar and gas-phase metallicity measurement of galaxies are still quite uncertain (e.g., \citealt{Goddard_2017} for stellar metallicity measurements with different stellar population synthesis codes and stellar templates and \citealt{Maiolino_Mannucci2019} for a review of the various methods for the gas-phase metallicity determination). This can also be seen in Fig. \ref{str_mzr_evo} and \ref{gas_mzr_evo}, where multiple sources of the metallicity measurements are included. It is very important to obtain the absolute values of metallicities of both stellar and gas-phase in observations. 
%The large systematic uncertainties of both stellar and gas-phase metallicity measurements may significantly weaken their constraining power on the supernova feedback mechanisms and thus deserve more attention. REPHRASE}

\subsubsection{Stellar MZR evolution}
\begin{figure*}
	\includegraphics[width=\textwidth]{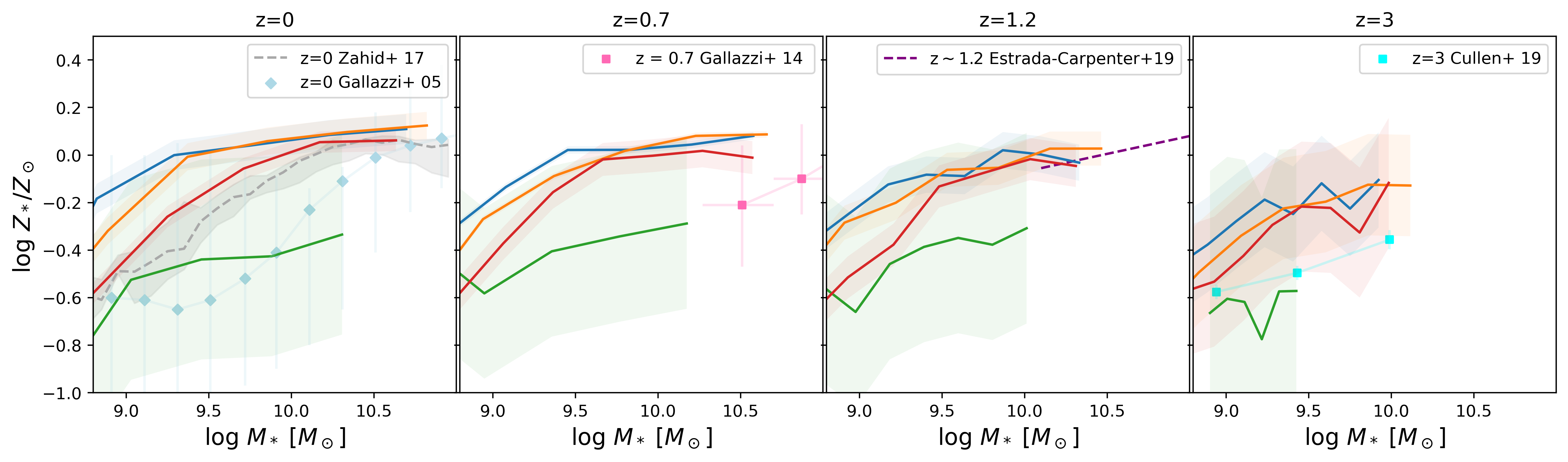}
    \caption{ Evolution of the luminosity-weighted stellar MZRs with thermal (blue), stochastic (orange), kinetic (green), and mechanical (red) feedback models. The solid lines are for the medians, and the shaded areas show the $1\sigma$ scatter. Observational data are taken from \citet[][$z=0$]{Zahid2017}, \citet[][$z=0$]{Gallazzi2005}, \citet[][$z=1.2$]{Gallazzi2014}, and \citet[][$z=3$, with $+0.5$ dex shift for {\rm [O/Fe]}]{Cullen2019}.
    %\ck{[Change all obs to dashed lines.For Choi+14, which data/element are you plotting? Change the 3rd panel to 1.3. Are you plotting the thick line of Fig.16 of their paper? In the 4th panel, plot Cullen+19 at z=3, shifted by +0.5dex for [O/Fe]]}
    }\label{str_mzr_evo}
\end{figure*}

\begin{figure*}
	\includegraphics[width=\textwidth]{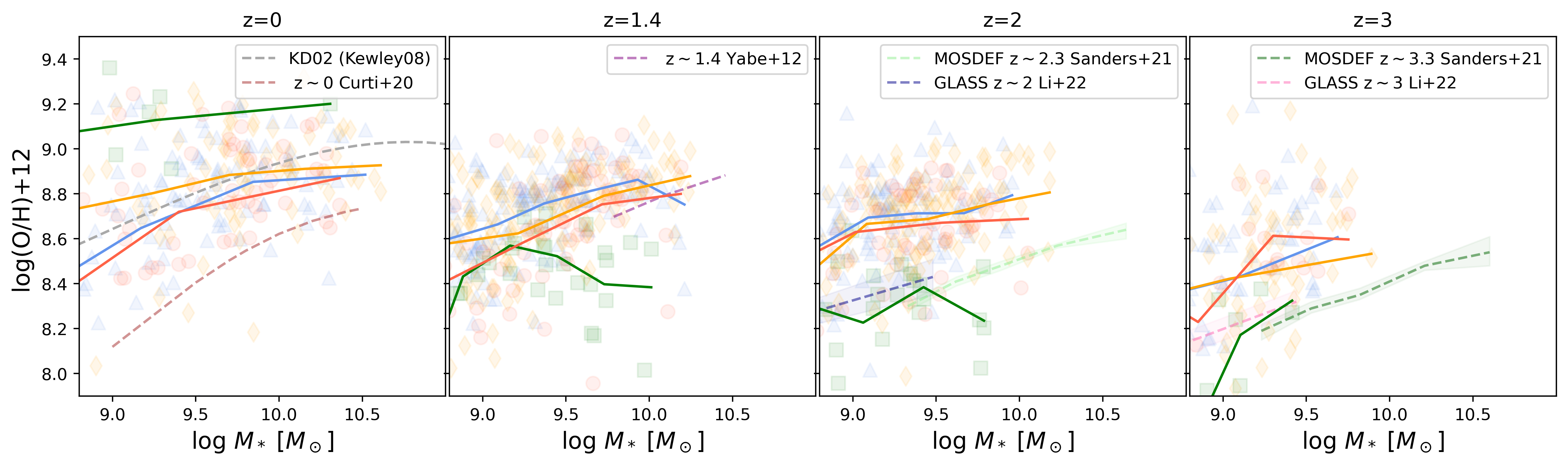}    \caption{ Evolution of the SFR-weighted gas-phase MZRs with thermal (blue; triangles), stochastic (orange; diamonds), kinetic (green;squares), and mechanical (red; circles) feedback models. The solid lines indicate the linear fit to the individual galaxies shown by the symbols with the same colour.
    Observational data are from \protect\citealt{Tremonti2004} ($z=0$) with the KD02 scale in \citet[][$z=0$]{Kewley2008}, \citet[][$z=0$]{Curti2020}, \citet[][$z=1.4$]{Yabe2012}, \citet[][$z\sim$ 2--3]{Sanders2021}, \citet[][$z\sim$ 2--3]{li2022}.}\label{gas_mzr_evo}
\end{figure*}

Figure \ref{str_mzr_evo} shows the stellar MZRs from $z=0$ to 3 for the four feedback models. %The solid lines show the medians of individual galaxies in our simulations, while the shaded areas show the $1\sigma$ scatter.
At higher redshifts, all models systematically give lower metallicities at a given mass, showing very similar differences among the models \di{($\sim$ 0.2 dex from $z=3$ to 0)}.
At all shown redshifts, the thermal feedback always produces slightly more metals than the other models. {The low cosmic SFR with the kinetic feedback results in significantly lower stellar metallicities than in the other models.} %while the kinetic feedback produces very few metals and no metals at all at $z=3$ \ck{Kinetic feedback is producing metals in Fig.7}. 
%Although in Fig.\ref{z_evolution} the kinetic feedback produces a small amount of metals at $z\sim3$, there is no galaxy in the plotted range at $z\ge3$.
At the low-mass end, the stochastic feedback gives metallicities slightly lower than the thermal feedback by $\sim$ 0.1 dex. Overall, the supernova feedback has a more significant impact on the metallicity at the low-mass end where low-mass galaxies eject more metals \citep{Kobayashi_2007}. 

The mechanical feedback seems to give the best match to the observations at $z=0$, although the observed stellar metallicities at higher redshifts are either lower or of galaxies with limited overlap in mass compared to the model prediction.
In the observations, massive galaxies have super-solar metallicities at $z=0$, which disappear at $z=3$.
At $z=0$, as already shown in Fig.\ref{stellar_MZR}, our model agrees well with the latest analysis by \citet{Zahid2017}, although these give significantly higher metallicities than in \citet{Gallazzi2005}. 
At $z=0.7$, although there is no overlap in the mass range, data from \cite{Gallazzi2014} is more consistent with our kinetic model. However, this data set does not reject the other models if we consider the significant offset between \citet{Zahid2017} and \citet{Gallazzi2005} data at $z=0$. We also note the large error bars of $\sim 0.2$ dex for these data.
Our mechanical model seems consistent with the Hubble Space Telescope observations at $z=$ 1.1--1.6 \citep{Estrada_Carpenter2019}. These data suggest that the mass--metallicity relation has not significantly evolved since $z\sim0$ at least at the massive end, contrary to 
%after $z<1$ as it is comparable with 
\cite{Gallazzi2014} at $z\sim0.7$.
At $z=3$, the UV observations from \citealt{Cullen2019} are for Fe abundances, are shifted by $+0.5$ dex taking account of [O/Fe], but still about $0.2$ dex lower than our predicted metallicities.

\begin{figure*}
	\includegraphics[width=\textwidth]{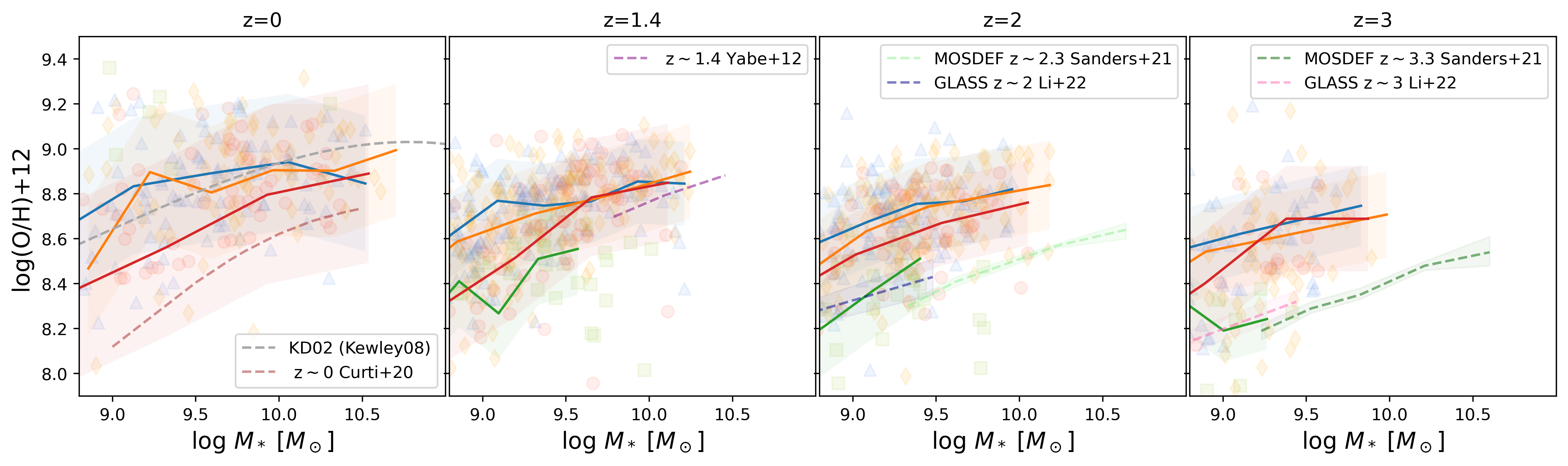}    \caption{ %\di{The symbols and the observational data (dashed lines) are the same as in Fig.\,\ref{gas_mzr_evo}.
 \di{Evolution of the young stellar MZRs (solid lines) comparing to the gas-phase metallicities (symbols, the same as in Fig.\,\ref{gas_mzr_evo}).} The solid lines are the medians of the luminosity-weighted stellar MZRs for stars younger than $0.1$ Gyr, and the shaded areas show the $1\sigma$ scatter.}  
 \label{gas_YoungStr_mzr}
\end{figure*}

\subsubsection{Gas phase MZR evolution}\label{sec_gas_ph_MZR_evo}
Figure \ref{gas_mzr_evo} shows the gas-phase MZRs from $z=0$ to 3 for the four feedback models.  %The scatter show the  individual galaxies in our simulations (triangle, diamond, square and circle for thermal, stochastic, kinetic and mechanical FB, respectively), while the solid lines show the median relations. 
\di{Although there is a significant scatter, and the sample is limited with our current resolution, there is a MZR at all redshifts.
There is also a redshift evolution from $z=$3 to 0, where} the metallicities decrease at higher redshifts \di{notably for the kinetic feedback ($\sim$ 1.4 dex). The thermal feedback has only a mild evolution ($\sim$ 0.05 dex), and the stochastic and mechanical feedbacks have a significant evolution ($\sim$ 0.2 dex). }
%The stellar MZR is lower for  compared to our other models 
%We find a large evolution from $z\sim3$ to 0 for the gas-phase MZR in the kinetic model \di{($\sim$ 1.4 dex), a very minor evolution for the thermal feedback ($\sim$ 0.05 dex), and a significant evolution ($\sim$ 0.2 dex) for the stochastic and mechanical feedbacks }(Fig.\ref{gas_mzr_evo}). 
%There is more significant evolution for the kinetic feedback compared to the other models. The metallicities are 0.8 dex higher at $z=0$ than at $z=3$. 
At high redshifts, most low-mass galaxies have fast star formation enriching their ISM. However, the kinetic feedback suppresses this star formation, which leads to only a few metal-poor low-mass galaxies. As discussed previously, the drastic change in the kinetic feedback model, which occurs exactly after $z=1$ removing metal-poor gas and causing the \di{metal-rich ($\sim$ 9.1 dex)} galaxies (green squares) at $z=0$.
%The evolution is less pronounced in the other models \di{($\sim$ 0.4 dex for the stochastic feedback, and insignificant for the thermal and mechanical feedbacks)}.

To verify the consistency between the stellar and gas-phase metallicities, we also show the metallicities of young stars since these are expected to be consistent with the metallicities of gas from which the stars were born. %the same look back time. 
In Figure \ref{gas_YoungStr_mzr}, we show the MZRs of young stars ($<$0.1 Gyr), comparing to the simulated gas-phase metallicities (\di{symbols}) and the observed gas-phase MZRs (dashed lines) at each redshift.
%The solid lines show the medians for the young stellar MZR, and the shaded areas show the 1 $\sigma$ scatter. 
The young stellar MZRs (solid lines) agree well with the simulated gas-phase metallicities. 
%With this method, however, we have better statistics, overlapping with the observations in a wider mass range. 
At $z=0$, the mechanical feedback increases with mass, similar to the observed MZRs. %and have a better match. 
At higher redshifts, the kinetic feedback fits better, as in Figure \ref{gas_mzr_evo}. Note that no line is plotted for the kinetic feedback at $z=0$ because not many young stars form with the kinetic feedback.
%To see when the stellar MZR is established, we included figure \ref{age} showing the age-unfolding for the stellar MZR at z=0.

The dashed lines represent observational data at various redshifts, which may be suffered by the uncertainties of the analysis methods, as already discussed. All observational data have been converted for the Kroupa IMF. At $z=0$, our mechanical feedback model most agrees with observed data from \protect\cite{Tremonti2004} with the KD02 scale in  \protect\cite{Kewley2008}, which gives \di{0.5--0.6} dex higher metallicities than in \protect\cite{Curti2020}.
At $z=1.4$, although we only have galaxies at the low-mass end, we compare with observational data from \citet{Yabe2012} (converted to the method from \citealt{Kewley2002} with the procedure given by \citealt{Kewley2008}) for massive galaxies. Then we find that the metallicity trend is comparable to our models.
At higher redshifts, the MOSDEF \protect\citep{Sanders2021} and GLASS survey with NIRISS slitless spectroscopy on the James Webb Space Telescope (JWST) \protect\citep{li2022} showed $\sim0.1$ and $\sim0.07$ dex evolution from $z\sim2$ to $z\sim3$, respectively, which is larger than in all our models ($\sim0.07$ dex), except for the kinetic.
The kinetic feedback model fits well with the laters, but as discussed previously, this model is underproducing stars, so this matching does not necessarily support kinetic feedback of supernovae.
%cannot be used for further predictions.}

%\ck{These could be reproduced with} the kinetic FB, but we know that our kinetic FB is underproducing stars \ck{compared with the observed cosmic SFR (Fig.\,6)}. Therefore, \ck{we cannot use this overlap for our predictions. [I know what you mean, but better way to say...]}

%%%%%%%%%%%%%%%%%%%%%%%%%%%%%%%%%%%%%%%%%%%%%%%%%%%%%%%%%%%%%%%%%%%%%%
%%%%%%%%%%%%%%%%%%%%%%%%%%%%%%%%%%%%%%%%%%%%%%%%%%%%%%%%%%%%%%%%%%%%%%
%\input{section4/conclusion}

\section{Conclusions}

Implementing four different methods of supernova feedback into our self-consistent cosmological chemodynamical simulations, we confirm that the modelling of feedback has a great impact on the mass--metallicity relations (MZRs), and can be constrained by spectroscopic observations of galaxies. In order to minimise other uncertainties, we have used the latest nucleosynthesis yields that can reproduce the observed elemental abundances of stars in the Milky Way \citep{Kobayashi_Karakas_Lugaro2020,Kobayashi_Leung_Nomoto2020}, and aim to reproduce the stellar and gas-phase metallicities simultaneously.

%In this paper, we study the impact of the modelling of supernova feedback on galaxy metallicities, using our self-consistent cosmological chemodynamical simulations that include the latest nucleosynthesis yields from \citet{Kobayashi_Leung_Nomoto2020,Kobayashi2020}. 
We compare four supernova feedback models: The classic thermal and kinetic models, where supernova energy is either ejected in pure thermal form or with a partial kinetic kick; The stochastic model, similar to \citet{Dalla2012}, which heats a random number of neighbour gas particles with a fixed energy increase; And the mechanical model from \citet{Hopkins2018}, which considers the work done during the Sedov-Taylor phase of supernova expansion.
After performing a parameter study (Appendix \ref{SFR_fiducial_f}),
%We study the four models with different values for the feedback parameter $f$ and 
we choose the following fiducial parameters from the observed cosmic SFRs (section \ref{fiducial}):  $f=1\%$, $f=50$, and $f=1\%$ for kinetic, stochastic, and mechanical feedback models, respectively.
Cosmic SFRs are significantly reduced with the kinetic feedback, which is too strong and is not producing enough stars, even with only a tiny fraction of supernova energy converted to a kick velocity. On the other hand, thermal and stochastic models are slightly overproducing stars at $z\gtsim2$. Mechanical feedback gives \di{a better} match to the observed cosmic SFRs (Fig. \ref{cosmic_SFR}). %model in our simulation as it compares best with observational data.

Despite fairly similar cosmic SFRs, we find a drastic change in the heating history of the ISM at $z\sim1$ with the kinetic feedback, and at $z\sim0.5$ with the stochastic feedback. This can be clearly seen in
the gas-phase space diagram (Figs. \ref{gas_pha_diag} and \ref{gas_pha_diag_z1}) as the hot diffused gas, as well as in the spatial distribution of temperatures (Fig. \ref{map_T}).
The spatial distribution of metals (Fig. \ref{map_Z}) are fairly similar, except for the kinetic feedback. %The stochastic feedback also causes a similar change but \di{at $z\sim0.5$.}
%\di{The gas-phase density temperature space diagram show the drastic behaviour of the kinetic model which overheats the diffused gas. }

Galaxy MZRs are greatly affected by the supernova feedback models. Strong supernova feedback makes star formation inefficient in the galaxy, which results in lower stellar metallicities of galaxies (Fig.\ref{stellar_MZR}). However, this is not the case for gas-phase metallicities, particularly with kinetic feedback (Fig.\ref{gas_mzr}). 
We find that young ($<0.1$ Gyr) stellar metallicities are consistent with the gas-phase metallicities.
Considering both stellar and gas-phase MZRs, our mechanical feedback %best matches
seems the most plausible in order to explain the observational data of present-day galaxies.

%We attempt a study of the stellar and gas-phase MZR evolution with the current resolution. 
Finally, we show the time evolution of the MZRs. As expected, both stellar and gas-phase metallicities become lower at higher redshifts in all feedback models. With our mechanical feedback, the predicted evolution of stellar MZR is in reasonably good agreement with the observations up to $z\sim3$ (Fig.\ref{str_mzr_evo}).
Our kinetic feedback model gives too low stellar metallicities at all redshifts.
%The stellar MZR is lower for  compared to our other models 
%We find a large evolution from $z\sim3$ to 0 for the gas-phase MZR in the kinetic model \di{($\sim$ 1.4 dex), a very minor evolution for the thermal feedback ($\sim$ 0.05 dex), and a significant evolution ($\sim$ 0.2 dex) for the stochastic and mechanical feedbacks }(Fig.\ref{gas_mzr_evo}). 
\di{For the gas-phase MZR, we find too large evolution in the kinetic model from $z\sim3$ to 0, and less prominent evolution for the other models} (Fig.\ref{gas_mzr_evo}). 
The available observations at $z>1$ seem rather consistent with the kinetic model, and we will investigate this further
%We aim to have a better estimation of the MZR evolution by running our simulation 
by comparing higher resolution and larger volume simulations to distant galaxies with the JWST as well as those of a large sample from ongoing and future spectroscopic galaxy surveys on ground-based telescopes.

%%%%%%%%%%%%%%%%%%%%%%%%%%%%%%%%%%%%%%%%%%%%%%%%%%%%%%%%%%%%%%%%%%%%%%

\section*{Acknowledgements}
We thank E. C. Lake, C. Lovell and J. Geach for fruitful discussions. We also thank the anonymous referee for useful comments. 
This work has made use of the University of Hertfordshire high-performance computing facility. 
This work also used the DiRAC@Durham facility managed by the Institute for Computational Cosmology on behalf of the STFC DiRAC HPC Facility (www.dirac.ac.uk). The equipment was funded by BEIS capital funding via STFC capital grants ST/P002293/1, ST/R002371/1 and ST/S002502/1, Durham University, and STFC operations grant ST/R000832/1. DiRAC is part of the National e-Infrastructure.
CK acknowledge funding from the UK Science and Technology Facility Council through grant ST/R000905/1, ST/V000632/1. 
The work was also funded by a Leverhulme Trust Research Project Grant on ``Birth of Elements''. 

%%%%%%%%%%%%%%%%%%%%%%%%%%%%%%%%%%%%%%%%%%%%%%%%%%
\section*{Data Availability}

The simulation data can be shared on request.

%%%%%%%%%%%%%%%%%%%% REFERENCES %%%%%%%%%%%%%%%%%%

% The best way to enter references is to use BibTeX:

\bibliographystyle{mnras}
\bibliography{ms}

% Alternatively you could enter them by hand, like this:
% This method is tedious and prone to error if you have lots of references
%\begin{thebibliography}{99}
%\bibitem[\protect\citeauthoryear{Author}{2012}]{Author2012}
%Author A.~N., 2013, Journal of Improbable Astronomy, 1, 1
%\bibitem[\protect\citeauthoryear{Others}{2013}]{Others2013}
%Others S., 2012, Journal of Interesting Stuff, 17, 198
%\end{thebibliography}

%%%%%%%%%%%%%%%%%%%%%%%%%%%%%%%%%%%%%%%%%%%%%%%%%%

%%%%%%%%%%%%%%%%% APPENDICES %%%%%%%%%%%%%%%%%%%%%
%\include{appendix}

\appendix

%%%%%%%%%%%%%%%%%%%%

%\section{Simulation}
%We employ a $\Lambda$CDM cosmology with $h$ = 0.68, $\Omega_m$ = 0.31, $\Omega_\Lambda$ = 0.69 and $\Omega_b$ = 0.048. The simulations are run at the same resolution with identical initial conditions: The same number of dark matter and gas with a resolution of  $N_\mathrm{gas}$=$N_\mathrm{DM}$=$128^3$.
%And in a periodic, comoving, cubic box volume of 10 $h^{-1}$Mpc.
%We generate star and gas particles as shown on the map in Fig. \ref{rho_maps}, using the Friend-of-Friends algorithm (FoF) to locate galaxies.

\section{Star formation rates}\label{SFR_fiducial_f}
\begin{figure*}
    \begin{subfigure}[]{0.45\textwidth}
    \includegraphics[width=1.1\textwidth]{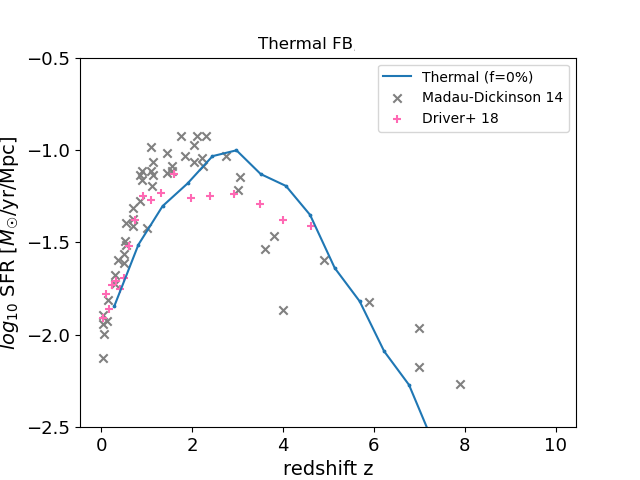}
    \caption{}\label{sfr_therm}
    \end{subfigure}
    \hfill
    \begin{subfigure}[]{0.45\textwidth}
    \includegraphics[width=1.1\textwidth]{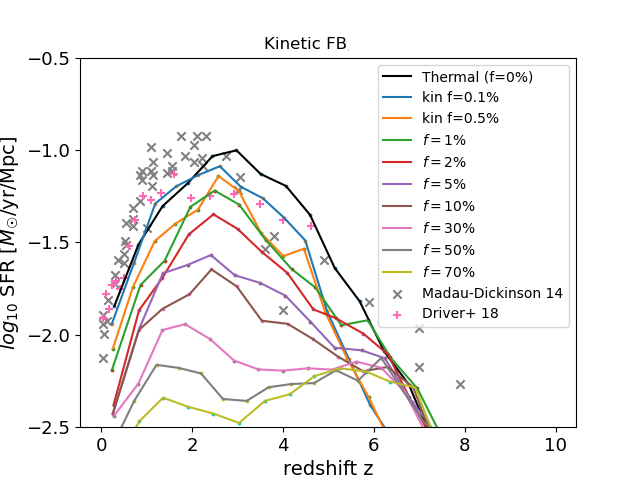}
    \caption{}\label{sfr_kin}
    \end{subfigure}
%\caption{common caption}
    \begin{subfigure}[]{0.45\textwidth}
    \includegraphics[width=1.1\textwidth]{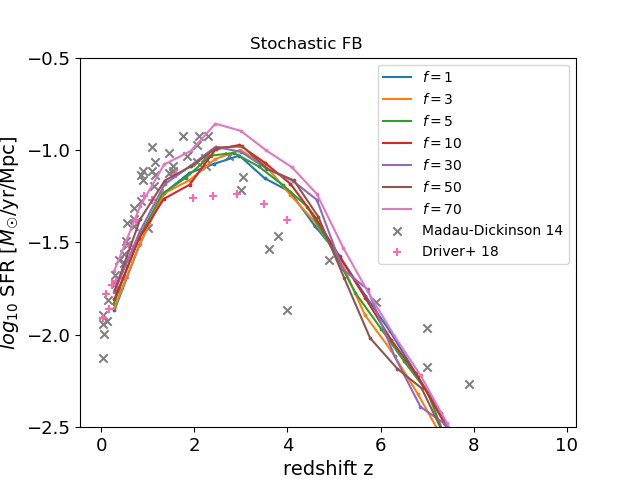}
    \caption{}\label{sfr_sto}
    \end{subfigure}
    \hfill
    \begin{subfigure}[]{0.45\textwidth}
    \includegraphics[width=1.1\textwidth]{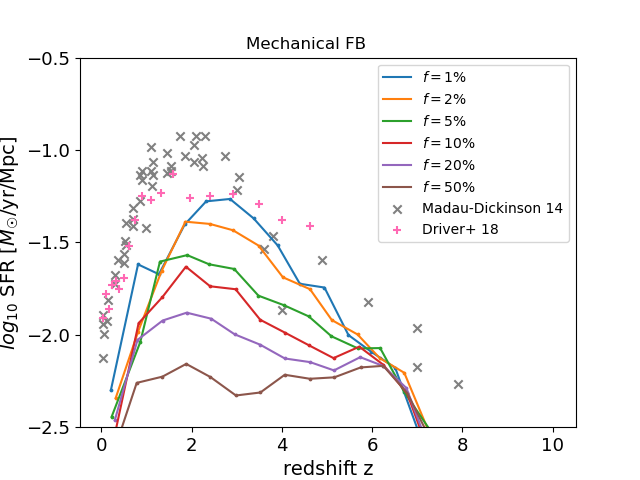}
    \caption{}\label{sfr_mec}
    \end{subfigure}
\caption{Cosmic SFRs for thermal, kinetic, stochastic and mechanical feedback models in panels (a), (b), (c) and (d) respectively. For each model we explore a wide range of feedback parameter $f$. The grey cross and magenta plus are observational data taken from \protect\citet{Madau_Dickinson2014} and \citet{Driver2018}, respectively, from far UV to mid IR.
}\label{sfr_allFB}
\end{figure*}

We have chosen the fiducial model parameters in order to match the observed cosmic SFR history.
Fig. \ref{sfr_allFB} shows the SFRs as a function of redshift for different values of the feedback parameter $f$ as obtained for thermal, kinetic, stochastic and mechanical feedback models, in panels (a),(b), (c), and (d), respectively. For this figure we use a resolution of $N_\mathrm{gas}$=$N_\mathrm{DM}$=$96^3$.
%The grey cross and magenta plus are observational data from \cite{Madau_Dickinson2014} and \citet{Driver2018}, respectively. 
All curves show a peak in the SFRs at $z \sim 3$. 
%except for mechanical feedback, where the peak is at $z \sim 2$, which is in a better agreement with observations. 
The box size in our simulation is limited due to the computation time. %limitations.
As a result, it does not include the formation of very massive galaxies and galaxy clusters at low redshifts. 
This explains the observed SFR peaks around $z \sim 2-3$, which are expected to be more consistent with observations for a larger simulation volume.
 For the kinetic and mechanical models, a larger $f$ results in a more efficient formation across cosmic time, but not for the stochastic model. The results of our parameter study can be summarised as follows.
\begin{itemize}
    \item Figure \ref{sfr_therm} shows the SFR for the thermal feedback. The SFR increases from $z\sim10$ to $z\sim3$,  
    %The peak at $z\sim3$ is reasonable with the cosmic noon period where star formation mostly occurred. 
    it decreases from $z\sim3$ to the present day because: (1) more gas has already turned into stars, (2) more supernova feedback suppressing star formation, and (3) more AGN feedback.
    \item Figure \ref{sfr_kin} shows the cosmic SFR with the kinetic feedback for different parameter values $f$. It shows that at high redshifts, the slope is the same for all parameters, as stars have not formed yet in these simulations. The feedback impact can only be seen after sufficient star formation has occurred, i.e. around redshift $z=6$. At redshift $z\leq6$, star formation is suppressed too much for $f > 30\%$. Then, the SFR slightly increases around $z=3$. This wave-shaped SFR history is explained independently of the feedback method by self-regulation: strong feedback suppresses star formation, resulting in less stellar feedback, which will, in return, increase star formation. (starting roughly at $z\sim4$ depending on the parameters). For a small parameter $f< 30\%$, the SFR increases from $z\sim6$ to $z\sim2$, where the feedback starts suppressing star formation.  The kinetic model with $f=0.1\%$ gives similar results to the thermal feedback. In order to demonstrate the impact of the kinetic part, we choose to use $f=1\%$ as our fiducial parameter. Overall, the kinetic feedback in our simulation is too strong and suppresses star formation too much, as even with $f=1\%$, the SFR peak remains too low compared to the observations.
    \item Figure \ref{sfr_sto} shows the cosmic SFR applying the stochastic feedback with different parameter values $f$. The SFR is larger for a larger $f$. This may be explained using Equation \ref{eq_De_sto} where the energy increase $\Delta e$ is proportional to $f$. Thus a large $f$ results in a large $\Delta e$, which yields the right-hand side of Equation \ref{eq_r_sto} to be small. Therefore, for a large $f$, Equation \ref{eq_r_sto} is rarely satisfied. Hence only a small number of particles receive the energy increase and are impacted by the feedback.
    When the condition is not satisfied, feedback does not impact the gas particles, which do not receive heating energy. The particles keep getting cool by following the cooling function until their temperature reaches $10^4$K. Once the particles are cool, the pressure is lost, the mater collapses toward the cooling particles where the density increases, and then the cooling rate becomes high (i.e. it accelerates the cooling). These features are shown in the star-forming region (low temperature, high density) of the gas-phase space diagram (Fig. \ref{gas_pha_diag}).
    \item Finally, mechanical feedback SFR is shown in Figure \ref{sfr_mec}, where we retrieve a similar behaviour as for the kinetic feedback, but slightly less efficient. We also find that this method is more affected by numerical resolutions than the other models, and have presented higher resolution results only in the previous sections.
\end{itemize}
For each method, we select the following fiducial parameters: $f=1\%$ (kinetic feedback), $f=50$ (stochastic feedback), and $f=1\%$ (mechanical feedback).

\section{Stellar Metallicities}
\begin{figure*}
    \begin{subfigure}[]{0.45\textwidth}
    \includegraphics[width=1.1\textwidth]{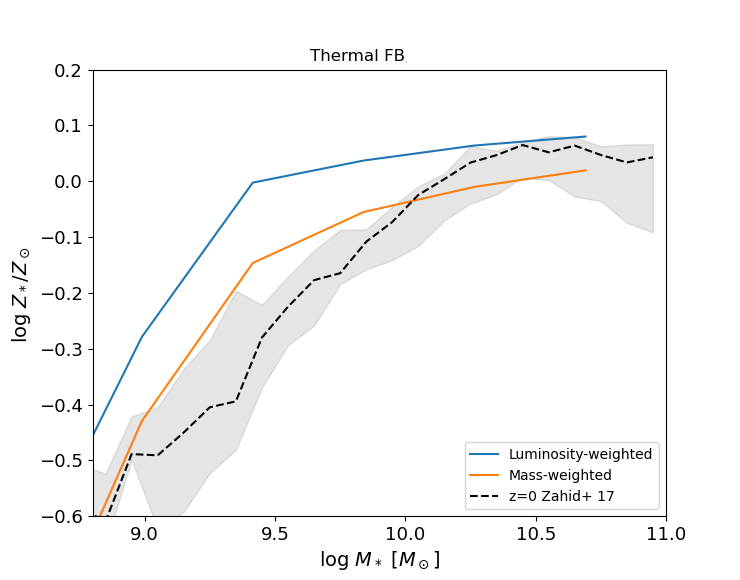}
    \caption{}\label{mzr_str_therm}
    \end{subfigure}
    \hfill
    \begin{subfigure}[]{0.45\textwidth}
    \includegraphics[width=1.1\textwidth]{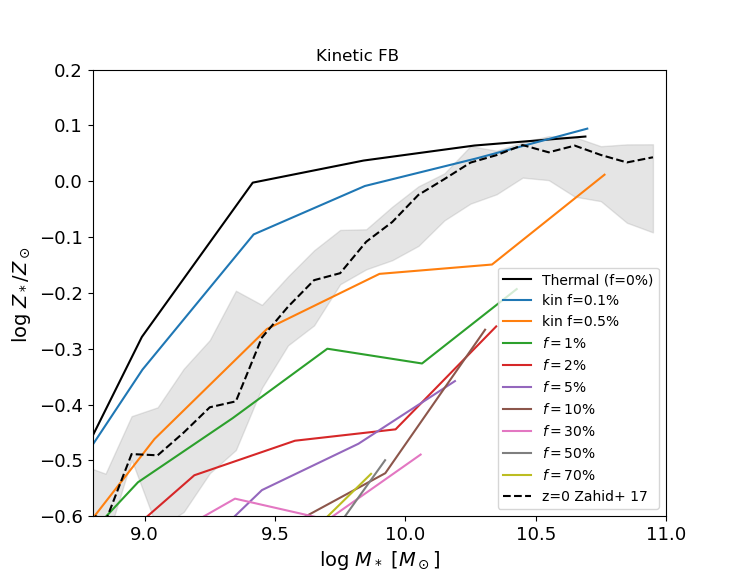}
    \caption{}\label{mzr_str_kin}
    \end{subfigure}
%\caption{common caption}
    \begin{subfigure}[]{0.45\textwidth}
    \includegraphics[width=1.1\textwidth]{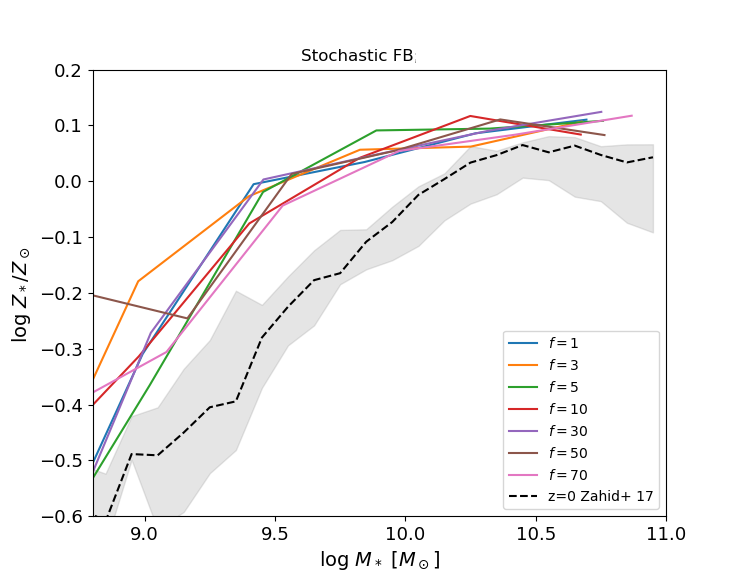}
    \caption{}\label{mzr_str_sto}
    \end{subfigure}
    \hfill
    \begin{subfigure}[]{0.45\textwidth}
    \includegraphics[width=1.1\textwidth]{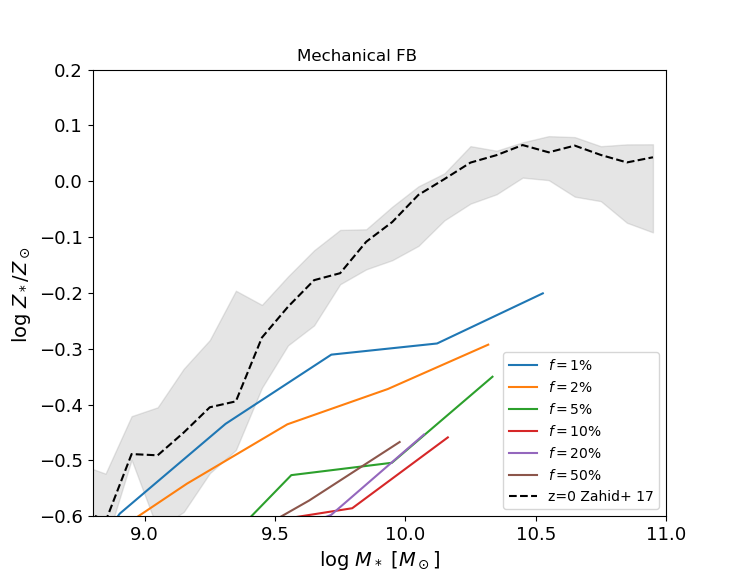}
    \caption{}\label{mzr_str_mec}
    \end{subfigure}
\caption{Stellar MZRs for thermal, kinetic, stochastic and mechanical feedback models are shown in panels (a), (b), (c) and (d), respectively. For the thermal feedback (panel (a)), we compare luminosity-weighted metallicity (blue) with mass-weighted metallicity (orange). For panels (b), (c) and (d), we explore a wide range of feedback parameters $f$, at $z=0$. The black dashed line and shaded are optical observational data taken from \citet{Zahid2017} of star-forming galaxies in the SDSS at $z=0$.}\label{mzr_str}
\end{figure*}

In what follows, we compare the impact of the feedback parameter $f$ on the stellar MZR at $z=0$ for each model.
\begin{itemize}
    \item The MZR for thermal feedback is shown in Figure \ref{mzr_str_therm}, comparing the luminosity-weighted metallicity (blue) with the mass-weighted metallicity (orange). There is a $0.2$dex offset; the luminosity-weighted metallicity is higher because it is weighted for young and metal-rich stars.
    \item Figure \ref{mzr_str_kin} shows the MZRs using the kinetic feedback model with different parameter values $f$. It shows that overall the metallicity is always lower than the observed MZR. The metallicity is lower for stronger kinetic feedback (larger $f$) because a large kinetic velocity ejects more outflows, driving the metal-enriched gas out of the galaxy. As explained above, this difference is more visible in low-mass galaxies.
    \item Figure \ref{mzr_str_sto} shows the MZRs with the stochastic feedback. At the high-mass end, the MZR is not impacted by the parameter $f$ and the metallicities are always higher than observed. Lower mass galaxies ($M<10^9 M_\odot$) have higher metallicities with a larger $f$. This agrees with what is discussed above for the SFR with the stochastic feedback where a larger $f$ produces more star formation, enhancing the metallicities.
    \item Finally, the mechanical feedback MZRs are shown in Figure \ref{mzr_str_mec} where we retrieve a similar behaviour as for the kinetic feedback. The metallicities are higher for a smaller $f$. 
\end{itemize}

\section{Age dependence of the stellar MZRs}
In Section \ref{sec_gas_ph_MZR_evo}, we have shown that young stellar MZRs are roughly consistent with gas-phase MZRs at $z=0$. Here, we show the MZR dependence on the age-unfolding at various redshifts, which might indicate when the stellar MZR is established.

%\di{Figure \ref{age} compares the stellar MZR with different stellar ages \ck{(ages of star particles)} for the four feedback models. \ck{A clear time evolution is seen; }old stars with 9--10 Gyr have lower metallicities since they formed in a less enriched environment. Young stars with 0.1--2 Gyr are more metal rich. Stars $< 0.1$ Gyr (blue line) looks less metal rich than green and orange lines because of the small sample at the massive end.}

\begin{figure*}
	\includegraphics[width=\textwidth]{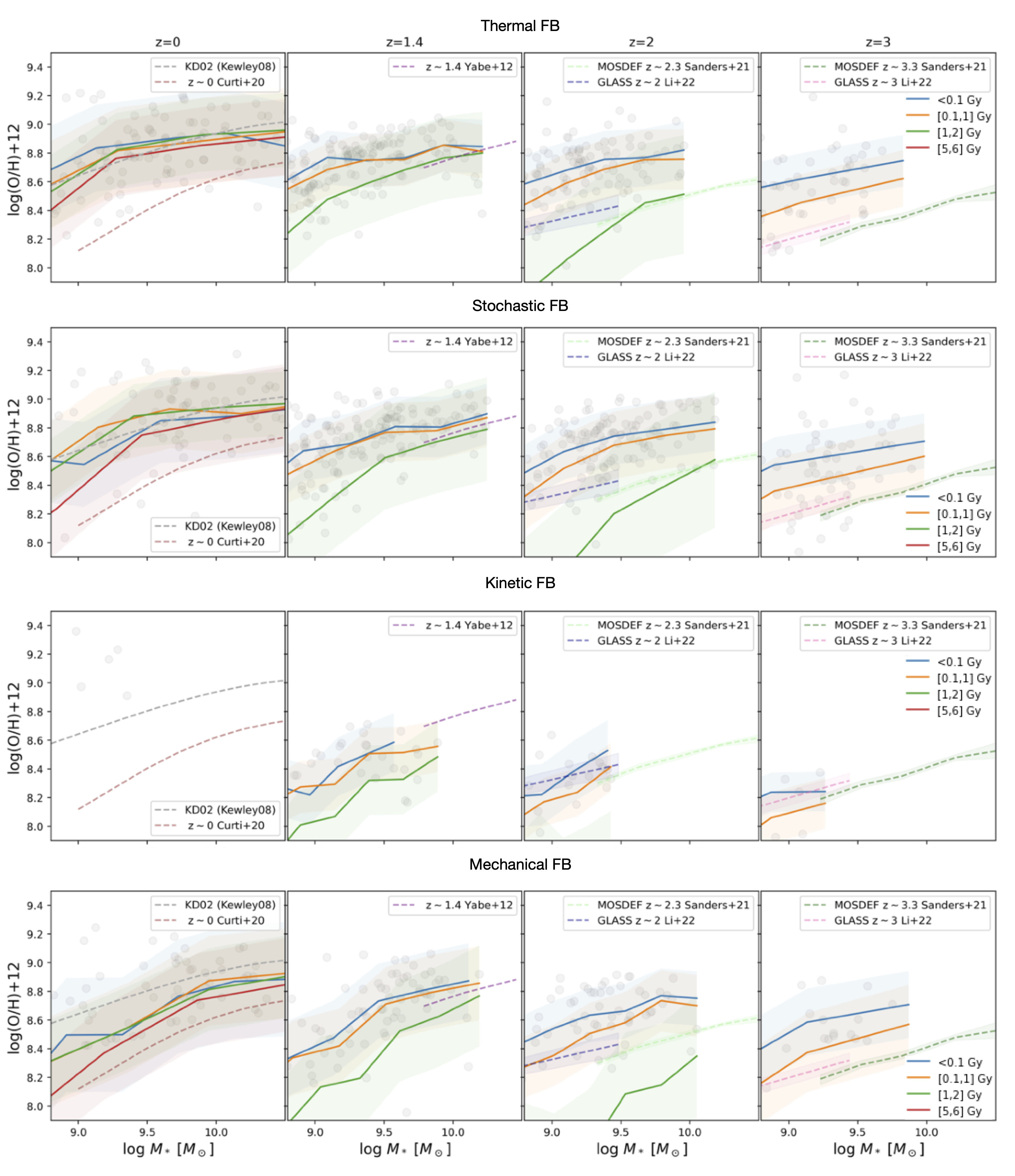}    \caption{Evolution of luminosity-weighted stellar MZRs of galaxies with different ages of star particles: $<0.1$ (blue), 0.1--1 (orange), 1--2 (green), and 5--6 Gyr (red), comparing to the gas-phase metallicities of galaxies (grey points). From top panel to bottom, the figure shows thermal, stochastic, kinetic and mechanical feedback models. The solid lines are for medians, and the shaded areas show the $1\sigma$ scatter. The dashed lines represent the same observational data for the gas-phase MZRs as in Fig.\,\ref{gas_mzr_evo}.}\label{age_z}
\end{figure*}

Figure \ref{age_z} compares the stellar MZRs with different stellar ages (ages of star particles) for the four feedback models.
A clear time evolution is seen; younger stars tend to have higher metallicities.
Stars $< 0.1$ Gyr (blue line) look less metal-rich than green and orange lines at the massive end because of the small sample. 
The MZRs of stars younger than $\sim$1 Gyr are consistent with simulated gas-phase metallicities (grey points) at all redshifts.
At $z\sim3$, MZRs with 0.1--1 Gyr old stars (orange line) show a similar slope as the other MZRs plotted in this figure; these stars have formed around $z\sim5$. % 2.140-1 = 1.140 Gyr => z=5.1
At $z\sim2$, MZRs with 1--2 Gyr old stars (green line) show significantly lower metallicities with a larger scatter; these stars have formed around $z\sim5$. % 3.272-2=1.272 Gyr => z=4.67
These might mean that MZRs are established at $z\sim5$, which might be consistent with the lack of a clear MZR in the recent JWST observations of $z\sim8$ galaxies \citep{Curti2023}.
Better statistics would be required to investigate this further.

%\begin{figure*}
%	\includegraphics[width=\textwidth]{AGE.png}    \caption{\di{Stellar MZRs for thermal, kinetic, stochastic and mechanical feedback models are shown in panels (a), (b), (c) and (d), respectively. We compare different bins of stellar ages \ck{using the ages of star particles: $<0.1$ Gyr (blue), 0.1--1 (orange), 1--2 (green), 5--6 (red), and 9--10 Gyr (purple)}. The black \ck{dashed [update the figure]} line and gray shade are \ck{the observed stellar MZR} taken from \citet{Zahid2017} %of star-forming galaxies in the SDSS 
% at $z=0$.}}\label{age}
%\end{figure*}

%%%%%%%%%%%%%%%%%%%%%%%%%%%%%%%%%%%%%%%%%%%%%%%%%%
%%%%%%%%%%%%%%%%%%%%%%%%%%%%%%%%%%%%%%%%%%%%%%%%%%
%%%%%%%%%%%%%%%%%%%%%%%%%%%%%%%%%%%%%%%%%%%%%%%%%%

% Don't change these lines
\bsp	% typesetting comment
\label{lastpage}
\end{document}